\documentclass[11pt,a4paper]{article}

\usepackage{amssymb}
\usepackage{color, graphics, graphicx}
\usepackage{xcolor,colortbl}
\usepackage{amsmath}
\usepackage{multirow}
\usepackage{fontenc}
\usepackage{lscape}
\usepackage{natbib} 
\usepackage{url}
\usepackage{bbm}
\usepackage{dsfont}
\usepackage{amsthm}
\usepackage{enumitem}
\usepackage{afterpage} 
\usepackage{longtable} 
\usepackage{booktabs} 
\usepackage{bbm}
\usepackage{flafter}
\usepackage{rotating}
\usepackage{tocbibind}
\usepackage[title, toc]{appendix}

\usepackage{algorithm}
\usepackage{algpseudocode}

\usepackage{caption}
\usepackage{subcaption}

\RequirePackage[colorlinks]{hyperref}
\hypersetup{
    colorlinks,%
    citecolor=blue,%
    filecolor=red,%
    linkcolor=blue,%
    urlcolor=blue
}

\usepackage{bm}
\usepackage{natbib}
\usepackage{amssymb}

\usepackage{mathrsfs}
\usepackage{xargs}
\usepackage{bbm}
\usepackage{booktabs} 
\usepackage{amsmath}

\usepackage{geometry}
\usepackage{xcolor,colortbl}
\newcommand{\modelName}{Additive Factor model}

\DeclareMathOperator*{\argmin}{arg\,min}

\newcommand{\R}{\mathbb{R}}

\newcommand{\E}{\mathbb{E}}

\newcommand{\bigO}{\mathcal{O}}
\newcommand{\bigOp}{\mathcal{O}_p}

\newcommand{\rank}{\pi}

\newcommand{\factor}{Z}
\newcommand{\Factor}{ {\bm{\factor}} }

\newcommand{\obs}{X}
\newcommand{\Obs}{X}
\newcommand{\idiosync}{\epsilon}
\newcommand{\func}{f}
\newcommand{\Rfunc}{g}
\newcommand{\RFunc}{\bm{g}}

\newcommand{\Func}{\bm{\func}}         

\newcommand{\compFunc}{h}       

\newcommand{\io}{i}       
\newcommand{\iv}{j}       
\newcommand{\il}{l}       

\newcommand{\iov}{ {\io,\iv} }
\newcommand{\ivl}{ {\iv,\il} }
\newcommand{\iol}{ {\io,\il} }

\newcommand{\setl}{ \il \in \{1, \dots, \nf \} }
\newcommand{\seti}{ \iv \in \{1, \dots, \nv \} }
\newcommand{\sett}{ \io \in \{1, \dots, \no \} }

\newcommand{\nf}{q}       
\newcommand{\no}{n}       
\newcommand{\nv}{p}       
\newcommand{\nov}{{\no \nv}}       

\newcommand{\spt}{\mathcal{H}} 

\newcommand{\svt}{\spt_{\no}}

\newcommand{\fracobs}{\frac{1}{\no} }
\newcommand{\fracvar}{\frac{1}{\nv} }
\newcommand{\sumobs}{\sum_{\io=1}^\no }
\newcommand{\sumvar}{\sum_{\iv=1}^\nv }
\newcommand{\sumfac}{\sum_{\il=1}^\nf }       
\newcommand{\sumbasis}{\sum_{k=1}^d }       
\newcommand{\meanobs}{\fracobs \sumobs }
\newcommand{\meanvar}{\fracvar \sumvar }
\newcommand{\sumObsVar}{\sumobs \sumvar }
\newcommand{\totalmean}{\frac{1}{\nov} \sumObsVar }

\newcommandx{\entropy}[3][1 = H, 3= \delta]{\mathscr{#1}(#3, #2) }
\newcommandx{\entropyBis}[3][1 = H, 3= \delta]{\mathscr{#1}\{ #3, #2\} }
\newcommandx{\entropyTer}[3][1 = H, 3= \delta]{\mathscr{#1}[ #3, #2] }
\newcommand{\norm}[1]{ \Vert #1 \Vert }    

\newcommand{\Cr}{\color{red}}
\newcommand{\Cb}{\color{black}}

\newcommand{\jb}[1]{\Cr #1 \Cb}

\newtheorem{theorem}{Theorem}
\newtheorem{lemma}{Lemma}

\newtheorem{proposition}{Proposition}

\newtheorem{model}{Model}      

\newtheorem{assumptionA}{Assumption}

\newtheorem{assumptionB}{Assumption}

\definecolor{oliverblue}{HTML}{006699}

\newcommand{\oyc}[1]{#1}

\title{Statistical Quantile Learning for\\
Large, Nonlinear, and Additive Latent Variable Models}
\usepackage{authblk} 
\author[1]{Julien Bodelet}
\author[2]{Guillaume Blanc}
\author[3]{Jiajun Shan}
\author[4,5]{\authorcr Graciela Muniz Terrera}
\author[1,6]{Oliver Y. Chén}

\affil[1]{Lausanne University Hospital, Lausanne, Switzerland}
\affil[2]{University of Zürich, Zürich, Switzerland}
\affil[3]{University of Geneva, Geneva, Switzerland}
\affil[4]{Ohio University, Athens, OH, USA}
\affil[5]{University of Edinburgh, Edinburgh, UK}
\affil[6]{University of Lausanne, Lausanne, Switzerland}

\date{}

\begin{document}

\maketitle

\begin{abstract}
The studies of large-scale, high-dimensional data in fields such as genomics and neuroscience have injected new insights into science. Yet, despite advances, they are confronting several challenges{, often simultaneously:} lack of interpretability, nonlinearity, slow computation, inconsistency and uncertain convergence, and small sample sizes compared to high feature dimensions. 
Here, we propose a relatively simple, scalable, and consistent nonlinear dimension reduction method that can potentially address these issues in unsupervised settings.
We call this method \textit{Statistical Quantile Learning} (SQL) because, methodologically, it leverages on a quantile approximation of the latent variables together with standard nonparametric techniques (sieve or penalyzed methods). 
We show that estimating the model simplifies into a convex assignment matching problem; we derive its asymptotic properties; we show that the model is identifiable under few conditions.
Compared to its linear competitors, SQL explains more variance, yields better separation and explanation, and delivers more accurate outcome prediction.
Compared to its nonlinear competitors, SQL shows considerable advantage in interpretability, ease of use and computations in large-dimensional settings.
Finally, we apply SQL to high-dimensional gene expression data (consisting of $20,263$ genes from $801$ subjects), where 
the proposed method identified latent factors predictive of five cancer types. The SQL package is available at \url{https://github.com/jbodelet/SQL}.

\textit{Keywords: High-dimensionality, Nonlinear model, Latent variable model, Generative models, Dimension reduction, GAN, VAE, Nonparametric estimation, Assignment Matching, Prediction.}
\end{abstract}

\section{Introduction}

Recent progress in science has witnessed
increasingly large and complex datasets. 
While larger and more complex datasets generally encode more information,  
they bring about unique statistical and scientific 
challenges.
First, these data are high-dimensional, oftentimes with dimensionality much larger than the sample size. If not accounted for, traditional statistical methods may result in non-unique solutions.
Second, these data are often nonlinear, not only between features, but also between features and outcomes. While linear models may still identify some effects, their estimates may be prone to bias 
or result in misidentification, and sometimes both.

A common way
to deal with the first issue is to use dimensionality reduction. 
By condensing and summarizing information into a small number of features, 
it effectively reduces data size, facilitates explanation and visualization, and, when estimation is concerned, may yield unique solutions. Such attractive properties, therefore, have interested machine learning scientists, biologists and engineers, not to mention statisticians. 
Traditional dimension reduction tools, such as Principal Component Analysis (PCA) and Factor Analysis (FA), however, assume linearity and therefore are adequate for features that are (at least approximately) linear.
For nonlinear data, these methods may fail, even with minor perturbations. For example, 
consider 
million of single nucleotide polymorphisms (SNPs)
in genomics research and hundreds of thousands of 
voxels (brain areas) in 
neuroimaging studies. 
The SNP-to-SNP
and voxel-to-voxel relationships are high-dimensional and nonlinear (\citealp{becht2019dimensionality}). Additionally, when (disease) outcome prediction is concerned, the relationship between the genes and the outcome and that between brain regions and the outcome are also high-dimensional and nonlinear.
The problem is further compounded when such data are usually only available in a few dozens or hundreds of subjects, a number that is much smaller than that of the features. 
Although progresses are already
being made (\citealp{lee2007nonlinear}, \citealp{van2008visualizing},
\citealp{mcinnes2018umap}) 
there is a pressing need for interpretable nonlinear dimension reduction and prediction techniques
for large-scale and high-dimensional data.

Nonlinear latent variable models, as they generalize factor analysis, have sparked the interest of both the statistical and the machine learning communities, in the form of nonlinear factor models in statistical science and (deep) generative models in machine learning.
Nonlinear factor models (see \citealp{Amemiya2002} and reference therein) usually rely on a parametric form.
The deep generative models, on the other hand, rely on  deep neural networks (DNN).
These latter are
known to be universal functional approximators (\citealp{hornik1989multilayer}) and 
belong to the class of sieve estimators although their asymptotic behavior are not fully understood (see \cite{shen2019asymptotic}). 
To estimate the deep generative models, there are two main general approaches: 
a variational approach, also known as variational autoencoders (VAE), see  \cite{Kingma2014}, and a simulation-based approach, called the Generative Adversarial Networks (GAN), see 
\cite{goodfellow2014generative}. 
These methods rely on a double DNN specification.
A first DNN model the conditional distribution of the features given latent variables (often called the ``generator''), while a second DNN is used to recover the latent space.
In VAE, an encoder function directly maps the data to 
the latent space 
by approximating the posterior distribution of the latent variables (see \citealp{Kingma2014} and \citealp{doersch2016tutorial} for a detailed review).
In GAN, one performs inference implicitly using a discriminator.

While both nonlinear factor models and deep generative models have advanced the analysis of high-dimensional nonlinear data, they are facing crucial methodological, and computational challenges. 
For nonlinear factors models, they
lack nonparametric inferences. 
For deep generative models, while they are
effective, especially given large sample sizes, and 
can potentially better handle the curse of dimensionality, to accurately recover 
the discriminator (e.g., in GAN) or the encoder (e.g., in VAE),
both nonparametric functions of the $p$-dimensional data, 
is prohibitively difficult for large dimensional and high-dimensional ($p>n$) scenarios (\citealp{poggio2017and}, \citealp{bauer2019deep}).
Additionally, deep learning methods suffer from several drawbacks,
preventing a wide acceptance in the statistical community. 
For example, the estimators are very sensitive to the choice of the hyperparameters and training suffer from issues such as the vanishing gradient problem and mode collapse.
From a practical point, the use of the double neural networks can sometimes make their training difficult, especially for GANs.
Even more importantly, the parameters of the generator, often of main interest in science, are not uniquely identified for deep generative models and therefore are hard to interpret: due to these challenges, the deep generative models are often referred to as ``black boxes''.

Here, to address these challenges, we introduce the \textit{Statistical Quantile Learning} ($\text{SQL}$), a new nonparametric estimation method, that 
\begin{enumerate}
    \item Deals with the general, dual problems of nonlinearity and the curse of dimensionality in large and high-dimensional nonlinear data analysis;
    \item Warrants identifiable and consistent estimates, with an asymptotic guarantee and fast convergence rate in large and high-dimensional settings.
    \item Overcomes the restriction of parametric assumptions in nonlinear factor models in statistics and the difficulty of hyperparameter specification and training in deep learning.
    \item Achieves better separation and explains more variability than linear competitors, such as PCA, in unsupervised learning.
    \item Outperforms VAE in large and high-dimensional settings and is competitive to VAE in low-dimensional settings.
\end{enumerate}

Before presenting the general method in Section \ref{sec. estimation} and its theory in Section \ref{sec. theoretical results}, here we first outline the model in relatively simple terms and summarise the attractive theoretical findings in words.
We consider $p$-dimensional data $\bm X$ and the model 
$\bm X = \bm f(\Factor ) + \bm\epsilon$,
where $\bm\epsilon$ is a random errors vector and
$\Factor \in \mathcal{Z}^q$ is a $\nf$-dimensional unobserved latent variable. 
The dimension $q$ is usually assumed to be much smaller than $p$  for dimension reduction purpose.
The unknown function $\bm f : \mathcal{Z}^q \rightarrow \R^p$ is called the generator. 
We restrict the generator to additive functions. This avoids the curse of dimensionality when the number of latent variables $q$ is large
and offers more interpretability.
The factors can come from some fixed distribution, usually the normal or uniform distribution. 

The key idea of the SQL method is to approximate the latent factors by their quantiles.
We show that finding these ``latent quantiles'' renders to solving an assignment matching problem.
The generator is estimated using  standard statistical nonparametric techniques, such as sieve and penalyzed methods.
This framework allows us to investigate the rates of convergence of the SQL estimates using empirical process theory (\citealp{VandeGeer2000}, \citealp{vandervaart2000asymptotic}) for both sieve and penalyzed estimators under weak conditions.
Critically, we find
that the rates of convergence improve as both the sample size and the dimension of the features increase. 
In particular, when the dimension $p$ is large, the SQL estimates reach the classical rates of convergence as if the latent factors were known and thus enjoy the ``blessing of dimensionality''.

The rest of the article is organized as follows.
In Section \ref{sec. estimation}, we introduce 
the model and the estimation method for SQL. 
In Section \ref{sec. theoretical results}, we discuss the theoretical properties of the model and of the estimators.
In Section \ref{sec. empirical}, we illustrate the finite sample performance of the estimators using simulation studies.
In Section \ref{sec. data analysis}, we apply SQL 
to high-dimensional gene expression data from cancer patients and use the extracted latent features to predict five cancer types. 

\section{Method }
\label{sec. estimation}

\subsection{The model}

We consider $p$-dimensional observations $\bm X_1, \dots, \bm \Obs_n \in \R^p$.
Suppose that the observations are associated to unobserved $q$-dimensional latent factors
$\bm Z_i=(Z_{i,1}, \dots, Z_{i,q}) \in \mathcal{Z}^q$
through an additive generator, i.e., $\func_j(\cdot) := \mu_j + \sumfac \func_\ivl(\cdot)$,
satisfying the following model:
\begin{equation}\label{eq. additive model}
\obs_\iov = \mu_\iv + \sumfac \func_\ivl(\factor_\iol) + \idiosync_\iov, 
\quad
i=1,\dots,n, \quad 
j=1,\dots,p
\end{equation}
where $\bm \epsilon_\iov$ are mutually independent random errors and independant of $\Factor_\io$.
Moreover, following the convention in the literature, 
we assume the identifiability condition that 
$$ \E(\func_{\ivl}(\factor_\iol) )=0, \text{for all } j=1,...,p \text{ and } l=1,\dots,q \oyc{.}$$
The assumption implies that $\E[X_{ij}]= \mu_j$.
Without loss of generality, we assume that the observations $X_\iov$ are centered and that $\mu_j=0$.
To build our estimation, we 
assume the following two
reasonable, but key assumptions on the latent factors and the generators.
\begin{assumptionA}(Assumption on the latent factors).\label{ass. rates mixing}
For each $l\in \{1,\dots,q\}$, the latent factors $\{\factor_\iol\}_{i=1}^n$ have marginal cumulative distribution function (CDF)
$ P_\factor^{(l)}:\mathcal{Z}\rightarrow [0,1]$. The CDF is invertible and admits a continuous derivative $p_Z^{(l)}(z)>0$ for all $z\in \mathcal{Z}$.
Moreover, the latent factors satisfy the ergodic property
$$
\sup_{z\in \mathcal{Z}} \left\vert \frac{1}{n}\sumobs \mathbbm{1}(Z_\iol\leq z ) - P_Z^{(l)}(z)\right| \leq Cn^{-1/2}
$$
with high probability for some constant $C>0$,
where $\mathbbm{1}$ denotes the indicator function.
\end{assumptionA}

Assumption \ref{ass. rates mixing} simply assumes that a Dvoretzky-Kiefer-Wolfowitz type inequality applies on the true factors. 
It is very general and allows the factors to be dependent as long as they are strictly stationary and ergodic.
For example, one may assume that they are $\phi$-mixing with $\sum_k \phi(k) < \infty$ (see e.g., \citealp{Kim1997}).
Note that we also allow dependence between factors.

Without loss of generality, throughout the paper we assume that the factors have the same marginal distribution, i.e., $P_Z^{(l)}= P_Z$.
In generative models, it is common to choose either the normal or the uniform distribution.
Moreover, we assume the following:
\begin{assumptionA}(Assumption on the functions).\label{ass. rates Lipschitz}
There exists $L_n>0$
that is not dependent on $p$, such that
$$ \max_{j\in \{1,\dots,p\}} \max_{l\in \{1,\dots,q\}} \vert \func_\ivl^0(z) -\func^0_\ivl(\tilde z )\vert \leq L_n \vert z - \tilde z \vert $$
for any $z, \tilde z \in [P_Z^{-1}(\frac{1}{n}), P_Z^{-1}(\frac{n}{n+1})]$ and satisfying 
$n^{-1/2}L_n \rightarrow 0$
as $n\rightarrow \infty$.
\end{assumptionA}
Assumption \ref{ass. rates Lipschitz} is a locally Lipschitz condition over all functions. It allows for non-globally 
Lipschitz functions if we adapt $L_n$.
For example, for function spaces where some of the $f_\ivl(z)$ behave as polynomials of order $k$, i.e., if there is an absolute constant $C$ such $\max_{j,l} f_\ivl(z) \leq  C z^k$, and when $P_Z=\Phi$ is the Gaussian CDF, then Assumption \ref{ass. rates Lipschitz} is satisfied for $L_n= Ck(\Phi^{-1}(1-n^{-1}))^{k-1}$.
Using $\Phi^{-1}(1-n^{-1}) \asymp \sqrt{\log(n)}$, we get $L_n=\bigO\left(\log(n)^{\frac{k-1}{2}}\right)$.

\subsection{Estimation}

Our estimation method is based on finding the ranks of the latent factors.
For each factor, consider the order statistics of the latent factors, defined by
$$
\factor_{(1),l} \leq \factor_{(2),l} \leq \dots \leq \factor_{(n),l}.
$$
With probability $1$, the inequalities are strict and there are no ties.
For each $l$, the rank statistics $\pi^{(l)}_1, \dots, \pi^{(l)}_n$ are permutations of $\{1,..., \no\}$ such that $\factor_{(\pi_i^{(l)})} = \factor_\iol$.
Throughout the paper, we will write the rank statistics, $\pi^{(l)}$, as elements of the set of permutations $\Pi_n $.
Using the 
delta method (see, e.g., \citealp{vandervaart2000asymptotic}), the order statistics can be approximated by the quantile function $P_Z^{-1}$. That is, for an integer $t$ such that $(t-1)/n< \alpha \leq t/n$, we have
$ \factor_{(t),l} = P_Z^{-1}(\alpha) + \bigOp(1/\sqrt{n})$.
Taking $t$ as $\pi_i^{(l)}$, we obtain
\begin{equation}\label{eq.S3 zstart-z is bigOp }
\factor_\iol = P_Z^{-1}\left(\frac{\rank_\io^{(l)}}{\no + 1}\right) +
\bigOp\left(\frac{1}{\sqrt{\no}}\right).
\end{equation}
Given the distribution of the factors  $P_Z$, 
it is sufficient to know the ranks $\pi^{(l)}$ to obtain an approximation of $Z_i$.
Then, after substituting, we have 
\begin{equation}\label{eq.S3 approx factors in function}
\func_\ivl(\factor_\io) \approx \func_\ivl\left(P_Z^{-1}\left(\frac{\rank_\io^{(l)}}{\no + 1}\right) \right).    
\end{equation}
We now show that the approximation in \eqref{eq.S3 approx factors in function} is accurate.
From the assumptions on $\func_\ivl$ and $P_Z$, the composite functions $\func_\ivl \circ P_Z^{-1}$ are continuous and belong to the space 
$\mathcal{G} \subseteq \{g:(0,1)\rightarrow \R, g \in C^{m}, \int_0^1 g(\xi) d\xi = 0\}$, where $C^m$ is the class of $m$-differentiable functions for some $m\geq 0$.
To build the estimation, we will make use of a suitable function class $\mathcal{G}_n$ that 
approximates 
or 
equals $\mathcal{G}$.
Given $\mathcal{G}$, we define the class of additive function by
$$
\mathcal{G}^{\oplus q} = \left\{g:(0,1)^q \rightarrow \R, g(\bm \xi )=\sumfac g_\il(\xi_l), g_\il \in \mathcal{G} \right\}.
$$
We also define $\mathcal{G}^{\oplus q}_n$ in a similar way.
In order to control the smoothness of the $\Rfunc_\iv$, we may use the smoothness penalty
\begin{equation}
    I^2(\RFunc)= \frac{1}{\nv} \sumvar\sumfac \int_0^1 \left(\Rfunc_\ivl^{(m)}(\xi)\right)^2 d\xi,
\end{equation}
where $\bm g$ denotes a $p$-dimensional vector of functions $(g_1, \dots, g_p)$ with $g_1,\dots, g_p\in \mathcal{G}^{\oplus q}$.
The following result motivates our estimation method.

\begin{lemma}[Approximation error]\label{lemma. approximation error}
Assume that $\{f_j\}_{j=1}^p$ and $\{\bm Z_i\}_{i=1}^n$ satisfy model \eqref{eq. additive model} and Assumptions \ref{ass. rates mixing} and \ref{ass. rates Lipschitz}.
Then, for $\lambda \geq 0$,
there are $\pi^{(1)*}, \dots, \pi^{(q)*} \in \Pi_n$ and $g^{*}_1, \dots, g^*_p\in \mathcal{G}_n^{\oplus q}$ satisfying
\begin{equation*}
\totalmean 
\left( 
\sumfac \Rfunc_\ivl^*\left(\frac{\pi^{(l)*}}{n+1}\right) -\func_\ivl(\factor_\iol)  \right)^2
+ \lambda I^2(\bm g^*)
= 
\bigOp\left(\frac{L_n^2}{n} + \tau^*_n\right)
,
\end{equation*}
where for $\xi_i:= \frac{i}{n+1}$, $i =1,\dots,n$, we have
$$
\tau^*_{n} := \min_{g_1,\dots,g_p\in \mathcal{G}_n^{\oplus q}} \totalmean \left(\sumfac g_\ivl\left(\xi_i\right)-f_\ivl\left(P_Z^{-1}(\xi_i)\right) \right)^2
+ \lambda I^2(\bm g).
$$
\end{lemma}
Note that $\tau^*_n$ is called the approximation error in the literature. The space $\mathcal{G}_n$ and the tuning parameter $\lambda$ will be chosen such that $\tau^*_{n} \rightarrow 0$. In the case that $\lambda =0$, $\tau^*_n$ is called the sieve error.
In other terms, Lemma \ref{lemma. approximation error} says that, for suitable class $\mathcal{G}_n$ and $\lambda$, one can estimate the additive model by obtaining estimators for $\Rfunc_\iv^* \in \mathcal{G}_n^{\oplus q}$ and $\pi^{(l)*} \in \Pi_n$.
In this paper, we thus focus on estimating the ranks, 
say $\hat{\pi}^{(l)}$, to construct an estimator of the latent variable, i.e., $\hat Z_\iol = P_Z^{-1}(\hat \pi_\io^{(l)}/(n+1))$.

We can now define the SQL loss function as
\begin{equation}\label{eq. loss function}
\begin{split}
\mathcal{L}(\bm\pi, \bm \RFunc) = \totalmean \left(\obs_{\iov} - \sumfac\Rfunc_\ivl\left( \frac{\rank_\io^{(l)}}{\no + 1}\right)\right)^2,
\end{split}
\end{equation}
where $\bm \pi := (\pi^{(1)}, \dots, \pi^{(\nf)})$ is the $q$-dimensional vector of permutations.
The SQL estimator is defined as the solution of
\begin{equation}\label{eq. estimators definition}
\begin{split}
(\hat{\bm \pi}, \hat{\bm g}) = \argmin_{\bm \pi, \bm g}\left\{ \mathcal{L}(\bm\pi, \bm \RFunc) + \lambda I^2(\bm g), \text{ for } \pi^{(1)},\dots, \pi^{(q)}  \in \Pi_n \text{ and } g_1, \dots, g_p\in \mathcal{G}_n^{\oplus q} \right\}
\end{split}
\end{equation}
where we introduce
a penalty term for some tuning parameter $\lambda \geq 0$ that is allowed to depend on both $n$ and $p$.
Given $(\hat{\bm \pi}, \hat{\bm g})$, we let the final estimates of the factors and generators be:
$$ 
\hat \factor_\iol := P_Z^{-1}\left(\frac{\hat \rank_\io^{(l)}}{\no + 1}\right)
\text{ and }
\hat \func_\ivl(\cdot) :=  \hat{g}_\ivl( P_Z(\cdot)).
$$
We define the estimator very broadly 
to make our framework relatively general and, therefore, able to handle many different types of
estimators.
Note that $\mathcal{G}_n$ may be chosen as the true function space and thus may not depend on $n$;
in this case, we 
write $\mathcal{G}_n = \mathcal{G}$.
When the function space depends on $n$, $\mathcal{G}_n$ will be chosen as a sieve space in order to guarantee convergence of the SQL estimator.
A sieve space for $\mathcal{G}$ is a sequence of approximating space $\{\mathcal{G}_n\}_{n=1}^\infty$,
where $\forall g^0 \in \mathcal{G}$, there exists $g_n \in \mathcal{G}_n$ such that $d(g_n, g^0) \rightarrow 0$ as $n \rightarrow \infty$
for a suitable pseudo-distance $d$. We refer to \cite{Grenander1981} and \cite{Chen2007} for a review on sieve estimators.
A particular case are 
nonpenalized sieve estimators where $\lambda = 0$.
In 
Section \ref{sec. theoretical results}, we will 
prove the rates of convergence for both sieve and 
penalized estimators.

\subsection{Computational Algorithm}

In this section, we describe  
the algorithm to optimize \eqref{eq. estimators definition}.
We start by introducing some matrix notation.
Denoting by $\bm R^*$ the vector $\left(\frac{1}{n+1}, \dots, \frac{n}{n+1}\right)^\top$ and for any $n$-dimensional vector $\bm R=(R_1, \dots, R_n)^\top$, we write 
$\bm G^{(l)}(\bm R)$ as
the 
$n\times p$-dimensional
matrix with rows $(g_{1,l}(R_i), \dots, g_{p,l}(R_i))$, for $i =1,\dots,n$. 
Denoting by $\bm P^{(l)}$ the permutation matrix corresponding to $\pi^{(l)}$, 
we can rewrite the loss function \eqref{eq. loss function} in matrix notation as
\begin{equation}\label{eq. loss function mat}
\mathcal{L}(\bm \pi, \bm g) = \frac{1}{\nov}\left\Vert \bm X - \sumfac \bm P^{(l)} \bm G^{(l)}(\bm{R}^*) \right\Vert_F^2,
\end{equation}
where $ \Vert \cdot \Vert_F$ is the Froebenuis norm.
It is relatively easy to see that \eqref{eq. loss function} and \eqref{eq. loss function mat} are equivalent, since permuting the elements of $\bm R^*$ is equivalent to permuting the rows of $\bm G(\bm{R}^*)$, so that $\bm G(\bm{P} \bm{R}^*)= \bm{P} \bm G(\bm{R}^*)$.

Central to the optimization problem is a backfitting algorithm.
Specifically, we start with intial values $\hat{\bm P}^{(1)}, \dots, \hat{\bm P}^{(\nf)}$, and $\hat{\bm G}^{(1)}, \dots, \hat{\bm G}^{(q)}$. For $l=1, \dots, q$, we compute the residuals
$$
\bm{U}^{(l)} = \bm\obs - \sum_{k \neq l} \hat{\bm P}^{(k)}\bm \hat{\bm G}^{(k)}(\bm{R}^*)
$$ 
and update $(\hat{\bm{P}}^{(l)}, \hat{\bm G}^{(l)})$ by solving
\begin{equation}\label{eq. estimator backfitting}
\min_{\bm{P}^{(l)}, \bm 
G^{(l)}} \frac{1}{np}\Vert \bm U^{(l)} - \bm P^{(l)} \bm G^{(l)}(\bm{R}^*) \Vert_F^2
     + \frac{\lambda}{\nv} \sumvar \int_0^1 (g_\ivl^{(m)}(\xi))^2 d\xi
\end{equation}
The cycle is then repeated until convergence.
Optimization of \eqref{eq. estimator backfitting} can be 
performed alternatively with respect to 
$\bm P^{(l)}$ and $\bm G^{(l)}$.
Importantly, given $\bm G^{(l)}$, optimization with respect to $\bm P^{(l)}$ is a linear assignment matching problem, which 
can be solved in $O(n^3)$ polynomial time with the Hungarian algorithm.
Although alternate optimization is convenient and generally works well in practice, 
there is, unfortunately, no guarantee of convergence.
To address this issue, we show, in the next section, that 
the problem can be solved jointly over $\bm P^{(l)}$ and $\bm G^{(l)}$, when the functional space is a linear sieve.

\subsection{Connection with the Quadratic Assignment Problem} 

Following the previous subsection,
we consider the case when the estimated functions $\Rfunc_\ivl$ that solve \eqref{eq. estimators definition} lie in a linear functional space.
There are two possible scenarios: 
(i) 
when $\mathcal{G}_n$ is taken as a linear sieve space;
(ii) when the optimization is 
conducted over the space of $\mathcal{G}_n = \mathcal{G}$ of twice differentiable functions.
Whereas scenario (i) is relatively straightforward,
to see (ii), we report the next proposition, which can be found in \cite{wahba1990spline} or \cite{gu2013smoothing}.

\begin{proposition} [Translation to natural splines] \label{prop. natural splines}
Let $\mathcal{G}$ be the Sobolev space of $m$-times continuously differentiable functions. Suppose $\mathcal{G}_n = \mathcal{G}$, $\lambda >0$ in \eqref{eq. estimators definition}, and assume that the minimizers $\hat \Rfunc_j \in \mathcal{G}^{\oplus q}$ of \eqref{eq. estimators definition} exist. Then the $\hat\Rfunc_\ivl$ are natural splines with knots at $\frac{i}{n+1}$, $i=1,\dots, n$.
\end{proposition}

Hence, Proposition \ref{prop. natural splines} shows that we can restrict to the space of natural splines instead of considering an infinite dimensional space.
It follows that, in scenarios (i) and (ii), the solutions are linear combinations of bases functions.

Define $\{\psi_1(\cdot), k= 1, 2,\ldots, d \}$, where $\psi_k:[0,1]\rightarrow \R$,
as a set of 
basis functions spanning the functional space $\mathcal{G}_n$.
Consider $d$ that grows
to infinity as $n\rightarrow \infty$. 
More specifically, in the case of 
penalization, one may choose $d$ as large as $n$, while in 
the sieve scenario, one
may choose $d<<n$.
Depending on the true parameter space of $g_\ivl$, one may consider different basis functions,
such as B-spline, wavelets, polynomial series, or Fourier series.
We thus parametrize $\mathcal{G}_n$ as
$$ 
\Rfunc_\ivl(\xi) =  \sum_{k=1}^d b_{\iv,k}^{(l)} \psi_k(\xi)
$$
for coefficients $b_{j,k}^{(l)}\in \R$.
Denote $\bm \Psi^*$ as the $n \times d$ dimensional matrix with rows $\left( \psi_1( \frac{i}{ n + 1 } ), \dots, \psi_d( \frac{i}{ n + 1 } ) \right) $
and $\bm B^{(l)} =(\bm b_{1}^{(l)}, \dots, \bm b_p^{(l)})$ as the $p\times d$ matrices 
consisting of coefficients 
$\bm b_j^{(l)}:= (b_{j,1}^{(l)}, \dots,b_{j,d}^{(l)} )^\top$.
The minimization problem \eqref{eq. estimator backfitting} becomes 
\begin{equation}\label{eq. estimator backfitting linear}
\min_{\bm{P}^{(l)}, \bm 
B^{(l)}} 
\frac{1}{np}\Vert \bm U^{(l)} - \bm P \bm\Psi^* \bm B^{(l)} \Vert_F^2
+ \frac{\lambda}{\nv} \sumvar \bm b_{j}^{(l)\top} \bm \Omega \bm b_{j}^{(l)}
\end{equation}
where $\bm \Omega$ is the $d\times d$ matrix containing the products of the second derivatives of the basis functions, that is $$ 
\Omega_{k,k'} = \int_0^1 \psi_k^{(m)}(\xi) \psi_{k'}^{(m)}(\xi) d\xi.
$$
Given $\bm{P}^{(l)}$, the above yields the 
penalized least squares estimates
$$
\bm B_\pi^{(l)} = (\bm\Psi^{*\top} \bm\Psi^* + \lambda \bm \Omega)^{-1} \bm \Psi^{*\top} \bm P^{(l)\top} \bm U^{(l)} 
$$
Note that the orthogonality of $\bm P^{(l)}$ implied $\bm P^{(l)\top} \bm P^{(l)}= \bm I_n$, where $\bm I_n$ is the identity matrix of dimension $n$.
Replacing $\bm B_\pi$ in \eqref{eq. estimator backfitting linear} gives 
$$
\min_{\bm P^{(l)}} 
\frac{1}{np}\Vert \bm U^{(l)} - \bm P^{(l)} \bm\Psi^* (\bm\Psi^{*\top} \bm\Psi^* + \lambda \bm\Omega )^{-1} \bm \Psi^{*\top} \bm P^{(l)\top} \bm U^{(l)} \Vert_F^2.
$$
Subsequently, let $\bm M_\lambda = \bm I -\bm\Psi^* (\bm\Psi^{*\top} \bm\Psi^* + \lambda \bm\Omega)^{-1} \bm \Psi^{*\top}$. 
We have
\begin{align*}
\Vert \bm P^{(l)} \bm M_\lambda \bm P^{(l)\top} \bm U^{(l)}\Vert_F^2
&= \text{trace}(\bm U^{(l)\top} \bm P \bm M_\lambda^\top \bm P^{(l)\top} 
\bm P^{(l)} \bm M_\lambda \bm P^{(l)^\top} \bm U^{(l)} ),\\
&= \text{trace}( \bm U^{(l)} \bm U^{(l)\top} \bm P^{(l)} \bm M_\lambda^\top \bm M_\lambda \bm P^{(l)\top} ),    
\end{align*}
where we use the cyclic property of the trace operator.
The 
minimization problem, therefore, reduces to:
\begin{equation}\label{eq. qap}
    \min_{\bm P}\text{trace}( \bm U^{(l)} \bm U^{(l)\top} \bm P \bm M_\lambda^\top \bm M_\lambda \bm P^\top ).
\end{equation}

Optimization of \eqref{eq. qap} is a well-studied quadratic assignment problem (QAP)
(see e.g., \citet{burkard1998quadratic}).
It is NP-hard and for $n > 30$, computing the exact solution is usually unfeasible.
Fortunately, one can find good approximate solutions and bounds. For example, the Gilmore Lawler algorithm finds the Gilmore Lawler bound (GLB) 
by solving a series of Linear Assignment Problems 
with $O(n^5)$ complexity. 
In this paper, to reduce the computational burden we find approximate solutions of \eqref{eq. qap} by using an iterated Hungarian algorithm.
That is we iterate over $k=1,2,\dots$
$$
\bm P^{(l)k+1} = \argmin_{\bm P}\text{trace}( \bm C_{k} \bm P^{\top} ).
$$
until convergence, where the cost matrix $\bm C_k= \bm U^{(l)} \bm U^{(l)\top} \bm P^{(l)k} \bm M_\lambda^\top \bm M_\lambda$ is updated at each iteration.
Note that an interesting property of problem \eqref{eq. qap} is that 
the solutions depend only on $n$ and does not depend on $p$.
This 
observation renders the treatment of this problem advantageous to address challenges in high-dimensional data analysis where $p$ is large and $n$ small.
It not only complements methods in machine learning that requires large $n$ and small $p$, but also brings promises to real world studies where there are a lot of features (e.g., hundreds of thousands of voxels in neuroimaging research and millions of single nucleotide polymorphisms (SNPs) in genetic studies) of interest but collecting samples (e.g., patients with rare but deadly genetic and brain diseases) is difficult or expensive.
We further confirm this property in our simulation studies (Section \ref{sec. empirical}).

\section{Theoretical properties}
\label{sec. theoretical results}
After describing the method, here we explore the theoretical properties of SQL. Specifically,
we first provide rates of convergence for the estimator introduced in Section \ref{sec. estimation} and then study the identifiability of the model.

\subsection{Asymptotic theory}

We begin with the study of the rates of convergence of the 
estimators.
Let $\func^0_\iv(\cdot) $, for $\seti$, and
$\factor_\io^0$, for $\sett$, be the true functions and factors, respectively, satisfying Model \eqref{eq. additive model}
as well as Assumptions \ref{ass. rates mixing} and \ref{ass. rates Lipschitz}. First, we present a few assumptions for
the asymptotic analysis.

\begin{assumptionA}[Subgaussian errors]\label{ass. rates subgaussian}
There are positive constants  $K$ and $\sigma$ such that 
$$ \max_{1\leq \iv \leq \nv, 1\leq \io \leq \no} K^2 \left( \E e^{\idiosync_{\iov}^2/K^2}-1 \right) \leq \sigma^2.$$
\end{assumptionA}

\begin{assumptionA}[Sieve]\label{ass. rates splines}
In the optimization \eqref{eq. estimators definition} the space $\mathcal{G}_n$ is spanned by basis functions $\psi_1, \psi_2, \dots, \psi_d$ such that
$$ 
\sup_{\xi \in (0,1) } \quad
\inf_{b_{1}, \dots, b_d\in \R }
\left| \sum_{k=1}^d \psi_k(\xi) b_k   - \func^0_\ivl(P^{-1}_Z(\xi))\right| = \bigO(d^{-\eta}),$$
as $d \rightarrow \infty$ and for some $\eta \geq 1$.
\end{assumptionA}

\begin{assumptionA}[Penalized estimator]\label{ass. Sobolev space}
For some $m\geq 0$, the space of function $\mathcal{G}$ is the Sobolev space 
$$
\mathcal{G} = \{g:(0,1)\rightarrow\R,g \in C^{m}, \int_0^1 (g^{(m)}(z))^2 dz < \infty\}
$$
\end{assumptionA}

The subgaussian condition in Assumption \ref{ass. rates subgaussian} is standard in empirical process theory, but can be relaxed at the cost of some other additional conditions (see, e.g., \citealp{VandeGeer2000}).
If $\func_\ivl^0(\cdot)$ belong\oyc{s} to a class of smooth functions, then Assumption \ref{ass. rates splines} is satisfied with standard basis such as B-splines, polynomials, or wavelets (see \citet{Chen2007} for a detailed discussion).

The following theorem provides the rate of convergence for
our estimation method in the sieve setting (i.e. without penalisation).

\newcommand{\totalRates}{ \frac{ \log \no }{\nv} + \frac{L_n^2}{n} +  \frac{d }{\no}   + 
\frac{1}{d^{2\eta}}
}
\newcommand{\totalOptimizedRates}{ \frac{\log n}{p}  + \frac{L_n^2}{n} + 
\frac{1}{\no^{2\eta/(1+ 2\eta)}}
}

\begin{theorem}\label{th. rates of convergence}
Under Assumptions \ref{ass. rates mixing}-\ref{ass. rates splines}, 
consider the general
optimization \eqref{eq. estimators definition} 
with $\lambda = 0$.
For $L_n^2=o(n)$ and $\log(\no)= o(\nv)$,
it holds that
\begin{equation*}
\totalmean\left(
\hat \func_\iv(\hat{\bm{\factor}}_\io)-\func_\iv^0(\bm{\factor}_\io^0)  \right)^2 =  \bigOp\left( \totalRates \right).
\end{equation*}
Moreover, selecting the sieve dimension as $ d \asymp \no^{1/(1+ 2\eta)}$ gives the optimal rates
$$
\frac{1}{\nov} \sumobs \sumvar \left(
\hat \func_\iv(\hat{\bm{\factor}}_\io)-\func_\iv^0(\bm{\factor}_\io^0)  \right)^2 =  \bigOp\left( \totalOptimizedRates \right).
$$
\end{theorem}

The rates of the penalyzed estimator also depends on $log(n)/p$ and $L_n^2/n$.
Recall that $L_n$ depends on both the distribution $P_Z$ and on the shape of the generator.
For example, for function spaces where some of the $f_\ivl(z)$ behave as polynomials of order $k$, and when the factors are normally distributed one can choose $L_n^2\asymp\log(n)^{k-1}$.
We now proceed to provide the rates for the penalized version.

\begin{theorem}\label{th. rates of convergence penalyzed}
Under Assumptions \ref{ass. rates mixing}-\ref{ass. rates subgaussian}
and \ref{ass. Sobolev space}
assume
$\log(\no)= o(\nv)$
and $L_n^2=o(n)$,
it holds that
\begin{equation*}
\totalmean\left(
\hat \func_\iv(\hat{\bm{\factor}}_\io)-\func_\iv^0(\bm{\factor}_\io^0)  \right)^2
+ \lambda I^2(\hat{\bm g})
=  \bigOp\left( 
\frac{ \log \no }{\nv} + 
\frac{L_n^2}{n} +
\lambda I^2(\bm g^0) +
\frac{\lambda^{-1/2m}}{n}+ \frac{\log(\lambda^{-1/2}\vee 1)}{n}
\right),
\end{equation*}
where we used the notation $a\vee b = \max(a,b)$.
Moreover, choosing $\lambda^{-1}\asymp (n I^2(\bm g^0))^{2m/(2m+1)}$ gives
\begin{equation*}
\totalmean\left(
\hat \func_\iv(\hat{\bm{\factor}}_\io)-\func_\iv^0(\bm{\factor}_\io^0)  \right)^2
=  \bigOp\left( 
\frac{ \log \no }{\nv} + 
\frac{L_n^2\vee 1}{n}+
\frac{I^{2/(2m+1)}(\bm g^0)}{
n^{2m/(2m+1)}
}
\right),
\end{equation*}
and $I(\hat{\bm g}) = O_p(1) I(\bm g^0)$.
\end{theorem}

The optimal tuning parameters $d$ and $\lambda$ may be selected using cross validation or through simulations.
Note that provided that $p$ is sufficiently large compared to $\log(n)$, and $L_n^2/n$ is close to $0$, we obtain the standard rates of convergence for sieve estimators as if the factors were known (see e.g. \citealp{vandegeer2000empirical}).
This means that SQL enjoys what the blessing of dimensionality.
The number of observations should not be too large with respect to the number of variables.
This point is confirmed in the numerical experiments in Section \ref{sec. empirical}. 
Indeed, in modern ``big data'' scenarios, despite the growing number of sample sizes such as those available in multi-center studies, the sample size generally does not grow as fast as the feature dimensionality (e.g., hundreds of thousands of voxels in neuroimaging research and millions of single nucleotide polymorphisms (SNPs) in genetic studies). For smaller consortium or individual projects, it is generally more difficult to collect large samples but relatively easy to obtain high-dimensional features. 
For example, it is relatively easy to obtain hundreds of thousands of cells from HIV patients who show strong immune responses to the virus but it is very difficult to collect such samples due, in part, to costs, and, in part, to the rarity of such patients. This is one of the reasons
why the SQL method may be particularly attractive to
high-dimensional data analysis with relatively small samples sizes.

\subsection{Identifiability}

Equally important to the asymptotic theory and rates of convergence is the identifiability. It ensures the uncovering of
interpretable factors and functions. To that end, here we
investigate the identifiability of the population model 
\begin{equation}\label{eq. population model}
\bm X = \bm f(\bm Z) + \bm \epsilon
\end{equation}
where, for clarity, we use the compact notation $\bm f = (f_1, f_2, \dots, f_p)$, $f_j(z)= \sumfac f_{j,l}$, and $\bm Z= (Z_1, \dots, Z_q)^\top$.

In latent variable models, it is well-known that the latent variables and the generator (or loadings in linear factor models) are not separately identifiable. For example, for
linear factor models, where $\bm f(\bm Z)= \bm \Lambda \bm Z$, we can always find $\tilde{\bm{\Lambda}} = \bm\Lambda \bm H$, $\tilde{\bm{Z}} =\bm H^{-1} \bm Z$, for any invertible matrix $\bm H$, and obtain the same distribution for the observations $\bm X$. To separately identify the factors and generators in linear models,
additional conditions are imposed (see e.g., \citealp{lawley1962factor}). The same logic applies for nonlinear models, but the assumptions required for identifiabiltiy are, as of yet, little investigated.

Here, we propose a set of assumptions that ensure the
identifiability for nonlinear additive factor models. We begin with some motivations.
In nonlinear models, 
one can take $\tilde{\bm Z} = \bm  H(\bm Z)$ and $\tilde{\bm f} = \bm f\circ \bm H^{-1}$ for any bijective map $H$, yielding $\Func(\Factor) = \tilde \Func(\tilde \Factor)$.
It is important to note that  $\tilde{\bm Z} = \bm H(\bm Z)$ and $\bm Z$ may have different distribution.
Imposing a specific distribution, usually the normal or uniform distribution, on the factors is,
therefore, the first key to obtain identifiability in latent variable models.
In this section, without loss of generality, we assume that the
factors follow a standard normal distribution, $Z_{\il}\sim \mathcal{N}(0, 1)$.
Furthermore, we make the following assumptions.

\begin{assumptionB}[Intrinsic dimensionality]\label{ass. identification intrinsic dimension}
The outcome variable $\bm Y=\Func(\Factor)$ has intrinsic dimension $q$, i.e., it exists no map $\bm h:\mathcal{U}\rightarrow \R^p$ and random variables $\bm U \in \mathcal{U}$ with dimension $dim(\mathcal{U})<q$ such that $\bm Y = \bm h(\bm U)$.
\end{assumptionB}

\begin{assumptionB}[Sufficient nonlinearity of the generator]\label{ass. identification sufficient nonlinearity}
The functions $f_\ivl$ are twice differentiable for all $z$.
Moreover, the $(\nv \times 2\nf)$-dimensional matrix $\bm \Delta(\bm z):= (\bm \Delta_1(\bm z), \ldots, \bm \Delta_\nv(\bm z) )^\top$, with rows\\
$\bm \Delta_\iv(\bm z)^\top := (\func_{\iv 1}'(z_1), \ldots, \func_{\iv \nf}'(z_\nf), \func_{\iv 1}''(z_1), \ldots, \func_{\iv \nf}''(z_\nf))$, is of full rank for any $\bm z = (z_1, \ldots, z_\nf)^\top \in \R^\nf$.
\end{assumptionB}

\begin{assumptionB}[Ordering of the factors]\label{ass. identification function ordering}
The following ordering is met: 
$$ \meanvar \E(\func_{\iv, 1}(\factor_{1}))^2 > 
\meanvar \E(\func_{\iv,2}(\factor_{2}))^2
> \ldots >
\meanvar \E(\func_{\iv,\nf}(\factor_{\nf}))^2.
$$
\end{assumptionB}

Assumption \ref{ass. identification intrinsic dimension} rules out a case where one of the factors could be perfectly explained by the others.
Assumption \ref{ass. identification sufficient nonlinearity} requires that $p \geq 2\times q$. It excludes the linear factor model as the second derivatives are zero for linear generators.
As in any factor model, it is possible to permute the order of the factors without changing the model.
To avoid this possibility, and without loss of generality, Assumption \ref{ass. identification function ordering} uniquely fixes the ordering of the factors.
As in in principal component analysis the factors are ordered by their contributions to the variability of the observations.
Note that the factors are allowed to be dependent.
The following theorem establishes the identifiability for the additive  factor model \eqref{eq. additive model}.

\begin{theorem}[Identifiability]\label{th. identification}
Under Assumption 
\ref{ass. identification intrinsic dimension}-\ref{ass. identification function ordering}, 
the factors in Model \eqref{eq. population model} are identifiable up to a sign.
Specifically, suppose there exist alternate $\tilde{\bm \func}$, 
and factors $\tilde{\bm\factor}$ satisfying $\Func(\bm Z) =\tilde\Func(\tilde{\bm Z})$, and both satisfying Assumption \ref{ass. identification intrinsic dimension}-\ref{ass. identification function ordering}, 
we have then $\tilde \factor_\il = \pm \factor_\il$ almost surely for $\setl$.
\end{theorem}

\section{Simulation experiments}
\label{sec. empirical}

To evaluate the general performance of the SQL method, we conduct a series of 
simulation studies under different combinations of sample sizes and dimensionalities, and compare the proposed SQL method against 
the Variational Autoencoder (VAE).
We choose VAE for comparison for two reasons. First, it is a state-of-the-art method for dealing with generative model\oyc{s.} Second, unlike GAN, the VAE 
provides an estimator of the factors (as SQL does) for comparisons.
We show that our method (1) improves, for fixed dimensionality, as sample size or dimensionality increases, (2) rivals with or modestly outperforms the VAE in low dimensional cases, and (3) considerably outperforms the VAE when the dimensionality is large.
In Section \ref{sec. data analysis}, we further show that the SQL method outperforms PCA in supervised learning.
We present the implementation details of the VAE in Appendix \ref{apx:vae-implementation}.

\begin{figure}[!h]\label{fig. functional boxplots}
    \centering\includegraphics[scale = 0.5]{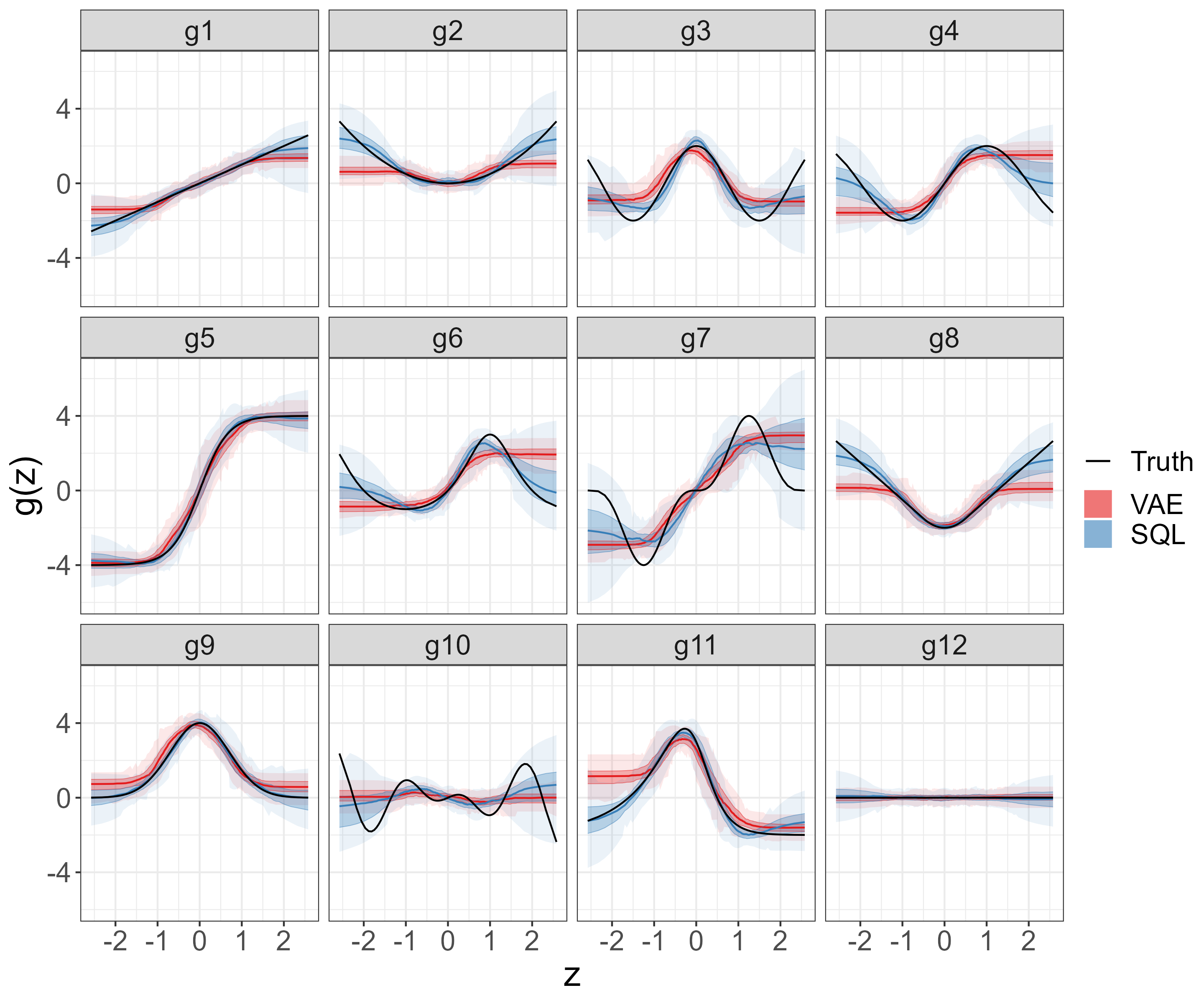}
    \caption{
    Functional boxplot of the $12$ functions composing the generator (Model \ref{Model 1}) when $n=100$. The black line represents the ground truth. The estimated functions of SQL are in blue. The estimated functions of the VAE are in red. We see that, across all 12 variables, the estimated functions using SQL are closer to the ground truth than those estimated by the VAE. This is observed for cases when the 
    ground truth is linear ($g_1$ and $g_{12}$), polynomial ($g_2$), trigonomotric ($g_3, g_4$, and $g_5$), and complex functions.}
    \label{fig. functional boxplots}
\end{figure}

\subsection{Simulation design}

We focus on two sets of 
common and general scenarios, low-dimensional and high-dimensional data, to investigate the properties of the estimators. 
In Model \ref{Model 1}, we 
investigate if SQL can challenge VAE in a classical setting,
that is, with a (very) low-dimensional model, with a small number of selected generators.
In this scenario, we focus on one factor and we let the sample size vary.
In Model \ref{Model 2}, we study an increasingly common type of data, that is, high-dimensional data, and investigate the impact of increasing dimensionality on the performance of the competing methods when the sample size is fixed. 
In this setting, we used randomly generated smooth functions to obtain the generator.

\begin{table}[h] \centering 
  \caption{
  Summary of the performance of the estimators in Model M1 ($p = 12 $). We report both the median and median absolute deviations (in parentheses) 
  over the $100$ 
 Monte Carlo runs.
  } 
  \label{tab. montecarlo_M1}
\begin{tabular}{@{\extracolsep{5pt}} ccccc} 
\\[-1.8ex]\hline 
\hline \\[-1.8ex] 
$n$ & \multicolumn{2}{c}{SQL} & \multicolumn{2}{c}{VAE} \\
 & $\text{mse}_z$ & $\text{mse}_g$ & $\text{mse}_z$& $\text{mse}_g$ \\ 
\hline
50 & 0.112 (0.071) & 0.46 (0.18) & 0.132 (0.057) & 0.516 (0.126)\\ 
100 & 0.095 (0.053) & 0.277  (0.1)& 0.118 (0.05) & 0.402 (0.106) \\ 
500 & 0.053  (0.015)& 0.087 (0.025) & 0.059 (0.039)& 0.172 (0.062)\\ 
1000 & 0.048  (0.009)& 0.059 (0.017)& 0.027  (0.004)& 0.079 (0.015)\\ 
\hline \\[-1.8ex] 
\end{tabular} 
\end{table}

\begin{model}[Low dimensionality]\label{Model 1}
We consider a framewwork with $q=1$ factor and $p=12$ variables, and we let $\no \in  \{50, 100, 500, 1000\}$. 
The $12$ functions are generated by
\begin{align*}
\func_{1}(\factor) &= \factor, \quad \qquad
&\func_{2}(\factor) &=\factor^2/2,  \\
\func_{3}(\factor) &=2\cos(\pi\factor/1.5),  \quad
&\func_{4}(\factor) &=2\sin(0.5\pi\factor), \\
\func_{5}(\factor)&= 4 \tanh(1.5\factor), \quad
&\func_{6}(\factor) &=\frac{3\sin(0.5\pi\factor)}{(2-\sin(0.5\pi \factor))}, \\
\func_{7}(\factor)&= 4 \sin(0.4 \pi Z)^3, \quad
&\func_{8}(\factor) &= 4 J(\factor)-2, \\
\func_{9}(\factor)&= 4\exp(-\factor^2),
& \func_{10}(\factor) &= Z \cos(3.5 Z), \\
\func_{11}(\factor)&= \frac{10 \exp( \factor )}{(1 + \exp\{4\factor\})}-2,
& \func_{12}(\factor) &= 0,
\end{align*}
where $J(Z) = 0.5\left(Z^2 \mathbbm{1}(|Z|<0.5) + (|Z|- 0.25) \mathbbm{1}(|Z|\geq 0.5)\right)$ is the Huber function with parameter $0.5$.
Both the factors and the errors 
are generated as independent $\mathcal{N}(0,1)$.
\end{model}
We present the results of the simulation study of Model \ref{Model 1} in Figure \ref{fig. functional boxplots} and Table \ref{tab. montecarlo_M1}.

\begin{model}[Large dimensionality]\label{Model 2}
We consider a framework where $\no = 200$ and
$\nv \in \{20, 50, 100, 200, 500\}$ and we consider $q\in \{1, 3\}$ latent factors.
To simulate the large number of functions, we generate 
them randomly using trigonometric functions. 
Specifically, we 
generate
$$
\tilde{f}_{j,l}(Z) = \frac{1}{C_{j,l}}\sum_{m=1}^4 \alpha_{j,l,m} \cos(2\pi m Z/ 8 ) + \beta_{j,l,m}\sin(2\pi m Z/ 8 ),
$$
where $\alpha_{j,l,m}$ and $\beta_{j,l,m}$ 
are both independently generated from $\mathcal{N}(0, m^{-2})$ for $j=1,...,p$, and $C_{j,l}$ are rescaling constants, where
$C_{j,l}= \sum_{m=1}^4 \alpha_{j,l,m}^2+ \beta_{j,l,m}^2$.
We then recenter the random functions, as
$$
f_{j,l}(Z) = \tilde{f}_{j,l}(Z) - 
\E[ \tilde{f}_{j,l}(Z) ]
$$
to ensure identifiability.
The factors $Z_{\iol}$ are 
generated as independent $\mathcal{N}(0,1)$ and the errors $\epsilon_\iov$ as $\mathcal{N}(0,1.5^2)$.
\end{model}
We present the simulation results of Model \ref{Model 2} in Table \ref{tab. montecarlo_M2}.

\subsection{Implementation}

To implement SQL, we use normalized B-splines (\citealp{schumaker2007spline}).
Given $M + 4$ evenly distributed knots, we obtain $M$ B-splines basis functions of order $3$, denoted as $\{\tilde{\Psi}_1, \tilde{\Psi}_2, ..., \tilde{\Psi}_M\}$.
As in \cite{boente2023robust},
we center the B-splines functions in order to guarantee that the estimated functions are also centered.
We thus define 
$\Psi_k(u) = \tilde{\Psi}_k(u)- \int_0^1 \tilde{\Psi}_k(u) du $ and use only the first $d= M-1$ basis functions to ensure that they are linearly independent.
We select 
$d=10$ for Model \ref{Model 1} and $d=8$ for Model \ref{Model 2}, respectively. Further, we select the
tuning parameter $\lambda$ 
using generalized 
cross-validation  
over a grid of length $20$.
We provide the details of the numerical implementation for the VAE in the Appendix \ref{apx:vae-implementation}.

\begin{table}[h] \centering 
  \caption{
    Summary of the performance of the estimators in Model \ref{Model 2} ($n = 200 $). We report both the median and median absolute deviations (in parentheses) 
  over the $100$ 
 Monte Carlo runs.
  } 
  \label{tab. montecarlo_M2} 
\begin{tabular}{@{\extracolsep{5pt}} cccccc} 
\\[-1.8ex]\hline 
\hline \\[-1.8ex] 
 & $p$ & \multicolumn{2}{c}{SQL} & \multicolumn{2}{c}{VAE} \\
 & & $mse_z$ & $mse_f$ & $mse_z$& $mse_f$ \\ 
$q=1$ & 20 & 0.922 (0.294) & 0.466 (0.126) & 0.866 (0.242) & 0.278 (0.104) \\ 
 & 50 & 0.34 (0.167) & 0.205 (0.052) & 0.512 (0.206) & 0.25 (0.067) \\ 
 & 100 & 0.11 (0.038) & 0.092 (0.021) & 0.245 (0.119) & 0.169 (0.034) \\ 
 & 200 & 0.049 (0.012) & 0.061 (0.005) & 0.105 (0.046) & 0.131 (0.020) \\ 
 & 500 & 0.033 (0.010) & 0.056 (0.004) & 0.102 (0.031) & 0.135 (0.014) \\ 
 \hline \\[-1.8ex]
$q=3$ & 20 & 1.003 (0.119) & 0.6 (0.056) & 0.924 (0.127) & 0.274 (0.049) \\ 
 & 50 & 0.846 (0.195) & 0.279 (0.048) & 0.937 (0.211) & 0.225 (0.039) \\ 
 & 100 & 0.433 (0.328) & 0.136 (0.083) & 0.796 (0.212) & 0.182 (0.037) \\ 
 & 200 & 0.059 (0.026) & 0.065 (0.005) & 0.748 (0.276) & 0.174 (0.047) \\ 
 & 500 & 0.031 (0.006) & 0.056 (0.001) & 0.723 (0.257) & 0.19 (0.057) \\ 
\end{tabular} 
\end{table}

\subsection{Simulation results}

For each set of parameters, we conduct 100 Monte Carlo experiments.
To evaluate the prediction performance of the estimation methods, 
we report the mean squared error of the factors, given by
$$
\text{mse}_z = \frac{1}{\nf\no}\sumobs \sumfac \left(\hat \factor_\iol - \factor_\iol \right)^2,
$$
as well as the mean squared prediction error of the generator, given by
$$
\text{mse}_f = \frac{1}{\nf \nv}\sumvar \sumfac \E\left[ \left(\hat \func_\ivl(\factor) -\func_\ivl(\factor)\right)^2\right].
$$
The above expectation is approximated by a sample of $1000$ points from the standard normal distribution.

In the classical setting of Model \ref{Model 1}, with $p$ fixed and small, the accuracy of both methods improve
as $n$ increases. SQL outperforms VAE for $n= 50, 100$, and $500$. 
For $n=1000$, however, the mean square error $\text{mse}_z$ 
is lower for VAE, while $\text{mse}_f$ 
is lower for SQL.
Figure \ref{fig. functional boxplots} depicts the estimated functions for both methods when 
$n= 100$. We can see that the SQL 
exceeds the VAE, especially on the tails.

In Model \ref{Model 2}, where $n$
is fixed, the performance of SQL 
improves significantly as $p$ increased. For VAE, there 
is a slight trend of improvement observed. SQL 
outperforms VAE when $p\geq 50$ for $q=1$ and when $p\geq 100$ for $q=3$. These simulations suggest three messages.
(1) In settings with small sample sizes and small dimensionality, SQL has a modest advantage over VAE. (2) In settings with large sample sizes and small dimensionality, SQL can compete with VAE. (3) When the dimensionality is larger, SQL has a significant advantage, and the advantage becomes increasingly obvious as the dimensionality grows.
These findings confirms empirically
the rates of convergence which states that the performance of SQL depends on $\log(n)/p$.

\section{Application to gene expression data}
\label{sec. data analysis}

After presenting the theory and methods and evaluating them via simulation studies, in
this section, we apply the SQL method to gene expression data of cancer patients. Many types of cancer have a genetic basis. The studies of the links between genes and oncological outcomes may not only help finding potential new genetic underpinings of specific cancer types but also help predicting the disease outcomes. Indeed, gene
expression data 
have been used to classify cancer (\citealp{golub1999molecular}). 
Yet, despite advances, most  cancer classifications rely on linear explorations. While linear methods are simple and easy to explain, this may naturally leave out a vast territory where genetic variables nonlinearly affect the outcomes. 
Additionally, when the dimensionality of genetic data is high, linear dimensional reduction may fail to uncover lower-dimensional representations whose relationship with the outcomes is non-linear. 
This problem may be further exacerbated when the lower-dimensional representations are used as features (predictors), as they perhaps only explain the portion of total disease variability that is linearly relevant.

The proposed SQL framework provides a platform to identify and estimate the lower-dimensional factors that are both linearly and nonlinearly related with, and therefore, better predict and potentially better explain, the outcomes.
In this paper, we apply SQL to investigate
the RNA-Seq dataset from the Pan-Cancer Atlas (PanCanAtlas) Initiative.
In brief, the PanCanAtlas data contain
gene expressions 
($p= 20,263$ genes with 
non-null expressions) 
from
$801$ patients.
Five types of tumors are present among these patients: breast cancer ($n=300$), kidney cancer ($n=146$), colon cancer ($n=78$), lung cancer ($n=141$), and prostate cancer ($n=136$).
See \url{http://archive.ics.uci.edu/ml} for detailed information about the dataset.

\begin{figure}[h]
    \centering\includegraphics[scale = 0.6]{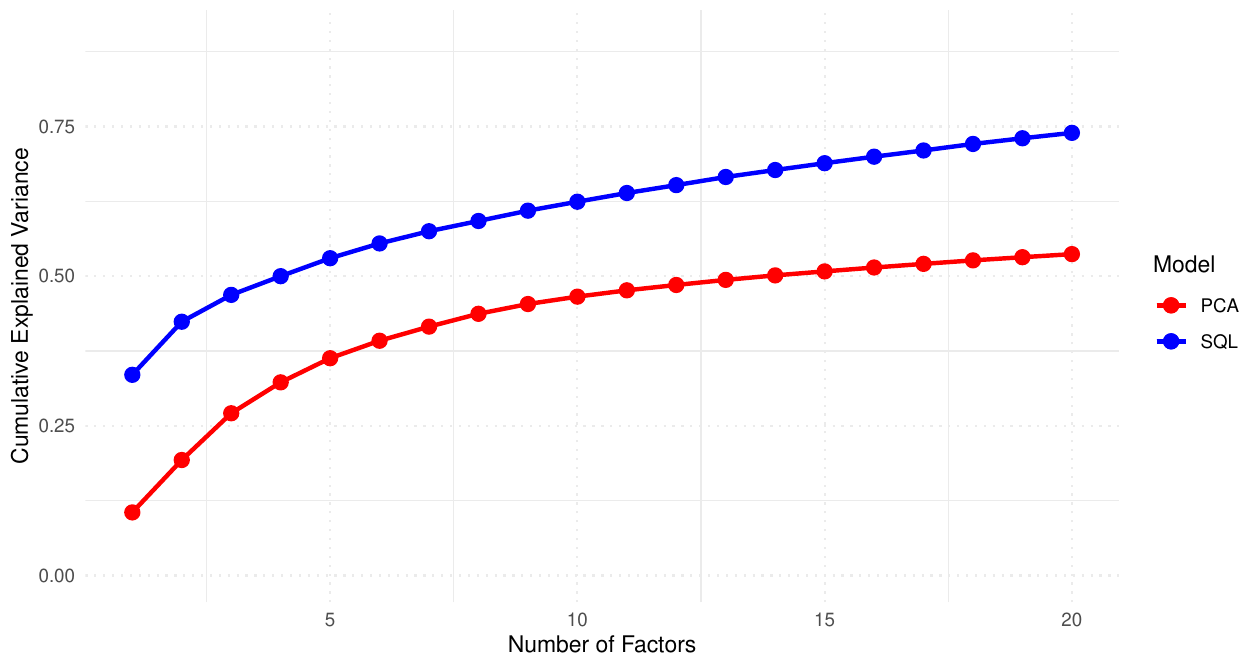}
    \caption{Proportion of the explained variance among total variance of the gene expression data with $p=20263$ genes. SQL explains $42.4\%$ of the variance with $2$ factors compared to $19.3\%$ by PCA. To explain $50\%$ of the variance, SQL requires only $4$ factors, while PCA requires $14$ components.}
    \label{fig. data analysis explained variance}
\end{figure}

We use SQL to perform cancer genetic data analysis in three domains: unsupervised learning, supervised learning, and latent space explainable analysis.

First, we apply SQL to 
study the data in
an unsupervised manner
to (a) find nonlinear latent factors that explain important variability of the total data; and (b) discover cancer-specific latent factors that separate cancer categories. We carry out SQL analysis vis-à-vis principal component analyis (PCA), a prominent linear competitor. 
Specifically, we begin by re-scaling the gene expression data.
Then, consistent to the steps
in the simulation study, we used normalized cubic B-splines bases,
with $d=12$.
We choose 
the Gaussian CDF as $P_Z$ and select the
tuning parameter 
using generalized cross-validation.
To select the number of factors, we 
use
\begin{equation}\label{eq. explained variance}
EV(q) = 1 - \frac{\sumobs \sumvar (X_\iov - \sumfac \hat{f}_\ivl(\hat\factor_\iol))^2}{
\sumobs \sumvar X_\iov^2
},
\end{equation} 
where the measure $EV(q)$ may be regarded as the proportion of the total variation explained by the model.
Finally, we
fit SQL for $q= 1,2,..., 20$, compute the proportion of explained variance, and compare it with the explained variance of PCA in Figure \ref{fig. data analysis explained variance}.
Our results suggest that, compared to PCA, SQL is able to explain the variance with fewer factors.
In particular, with $2$ factors, SQL explains $42.4\%$ of the variance compared to $19.3\%$ by PCA.
To explain $50\%$ of the variance, SQL needs only $4$ factors, while PCA needs $14$ factors.

\begin{figure}[h]
    \centering
    \begin{subfigure}{0.49\textwidth}
        \centering
        \includegraphics[viewport=0 30 350 350,clip, scale=0.52]{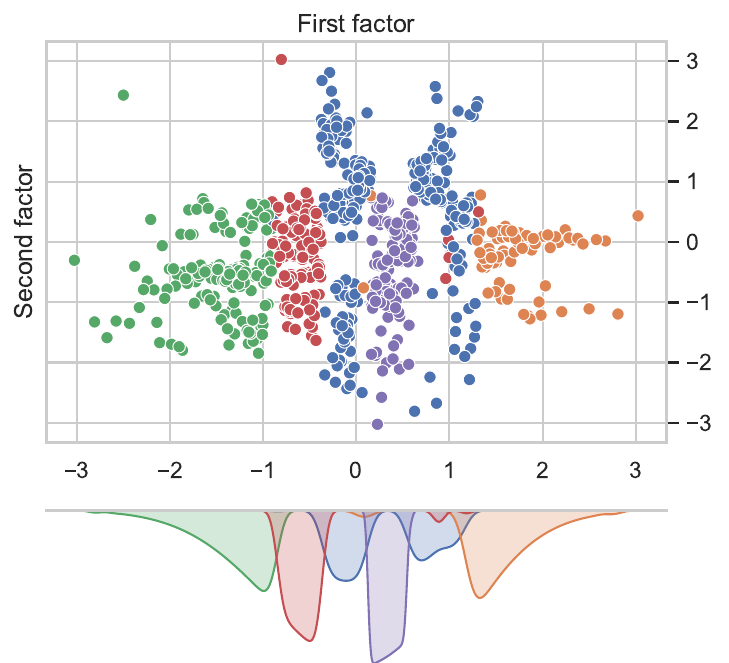}
        \caption{SQL}
        \label{fig. latent space sql}
    \end{subfigure}%
    \hfill
    \begin{subfigure}{0.49\textwidth}
        \centering
        \includegraphics[viewport=350 30 800 350,clip, scale=0.52]{figures/plot_latent_space.pdf}
        \caption{PCA}
        \label{fig. latent space pca}
    \end{subfigure}
    \caption{Visualization of the two-dimensional latent space of the 20,263 genes from 
    $801$ patients. The embedding plots suggest that SQL 
yields a better separation of five cancer types along its first factor (Panel a) compared to PCA along its first component (Panel b). 
The improved separation via SQL is
further illustrated by the conditional density estimates depicted at the bottom of each plot. The colour codes for the dots and histograms are blue = breast cancer (BRCA), yellow = colon cancer (COAD), green = kidney cancer (KIRC), red = lung cancer (LUAD), and purple = prostate cancer (PRAD).}
    \label{fig. latent space}
\end{figure}

Second, we use the identified latent features to classify cancer types using supervised learning. To that end, we feed the first two latent factors from SQL into the Support Vector Machine (SVM) to classify five cancer types.
To avoid a chance fitting (a good fit due to a lucky training/test split), we implement 
repeated random sub-sampling validation, with test sets of size $160$ and $100$ random sub samples.
This yields a
mean accuracy of $97.16\%$ (sd$=1.32$) for SQL, compared to 
$69.87\%$ (sd$=3.12$) 
using PCA.
Next, we enquire into the potential reason why SQL outperforms PCA. To do so, we plot
the first two factors against each other for 
SQL and PCA, respectively, in Figure \ref{fig. latent space}.
We can see that, graphically, 
SQL yields better 
(nonlinear) separation of five cancer types compared to PCA. This suggests that SQL is able to capture the nonlinearity in the data while PCA does not do as well.

\begin{figure}[h!]
\centering
\includegraphics[scale = 0.5]{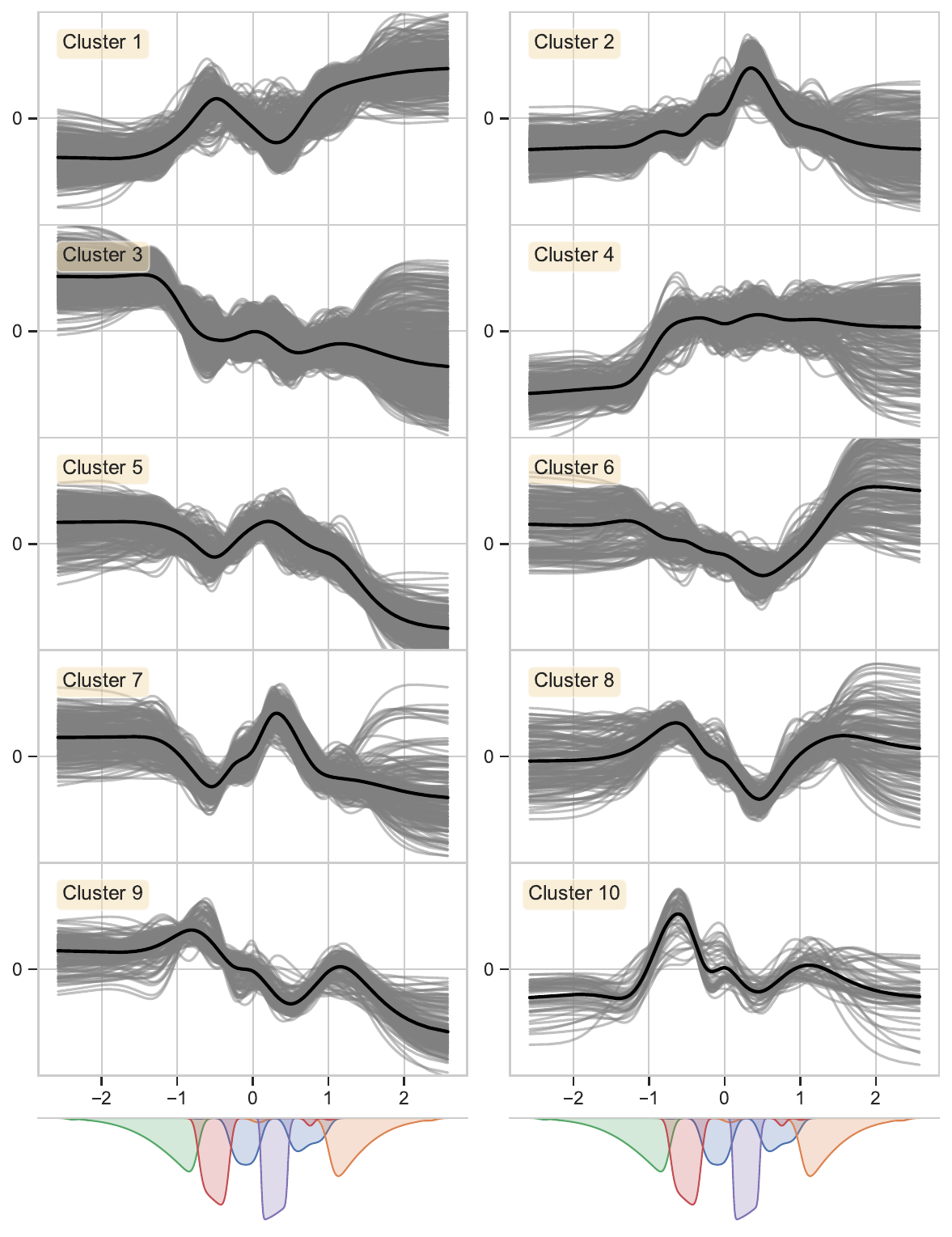}

\caption{Visualization of $3,000$ generators 
that best explain the data 
via $10$ functional clusters. To help interpretation, we align the histograms
of the factor for each cancer type 
on the x-axis. The colour codes for the histograms are blue = breast cancer (BRCA), yellow = colon cancer (COAD), green = kidney cancer (KIRC), red = lung cancer (LUAD), and purple = prostate cancer (PRAD).
The results show that genes corresponding to  
functional cluster 4, for example, play a significant role in discriminating kidney cancer.
}
\label{fig. functional clustering}
\end{figure}

Lastly, we investigate the relationship between the first
extracted latent factor and the gene expression features. 
As visualizing 
$20,263$
functions 
is difficult and potentially not meaningful, we 
first rank the functions and then select  
the top functions that best explain the data.
One effective way to rank the functions are sorting them according to their norms
$\Vert g\Vert_\varphi^2 = \int_{-\infty}^\infty g(z)^2 \varphi(Z) dz$, where $\varphi$ is the Gaussian density.
We then 
select the 
top $3,000$ functions with the highest norms and fine-groom them using  
hierarchical clustering  with distance metric
$d(g_1, g_2) = \Vert g_1-g_2 \Vert_\varphi$.
This yields $10$ clusters, as plotted in Figure \ref{fig. functional clustering}. In order to explore how genes relate to the cancer types via these functions, we add histograms
of the factors for each cancer type along
the x-axis, with which one can better interprete the clusters.
For example,
genes corresponding to 
functional cluster 4 play a significant role in discriminating kidney cancer
from other cancer types.
For genetic pathologists, it may be of further interest
to explore the roles of the genes
in each cluster and how they may give rise to the oncological outcomes via individual clusters using, for example, functional pathway analysis. 
This is, however, beyond the scope of the present paper and we will leave it to future work.

Taken together, using high-dimensional genetic data from patients of five types of cancer, we show
that the SQL method is able to uncover group-specific factors that not only better separate the groups but also explain a larger amount of variability than their linear counterparts. Additionally, using the latent factors as predictors, SQL is able to yield better out-of-sample prediction performance (than linear competitors) to classify cancer types in previously unseen subjects. Finally, the latent factors, as well as their associated functional clusters, help deliver meaningful scientific interpretation regarding the lower-dimensional representations and potentially identify and isolate functional genetic clusters related to cancerous outcomes.

\section{Conclusions}

In this paper, we introduce a new statistical method, \textit{Statistical Quantile Learning} ($\text{SQL}$), to study large-scale nonlinear high-dimensional data. 
Methodologically, by using a quantile approximation, SQL incorporates the characteristics of nonparametric statistics and constitute an alternative to deep generative models, overcoming some of their limitations. 
Compared to nonlinear factor models, SQL flexibly takes advantage of the rich nonparametric space. 
Theoretically, SQL is identifiable and the rates of convergence improve as both the sample size and feature dimensionality increase. 
Empirically, our simulations suggest that SQL competes with VAE in settings with relatively large sample sizes, and has a significant advantage in large and high dimensional settings. 
Additionally, a study of high-dimensional gene expression data from cancer patients shows that SQL extends even to supervised learning: the extracted latent factors, predictive of five types of cancer, can potentially serve as biomarkers.

There are a few directions this paper has not explored. 
First, we consider additive functions for the generator because it avoids the curse of dimensionality and provides more avenues for interpretability. 
However, this omits the scenarios where the generator is a multivariate function of the factors. {Future work may investigate this and extend our framework to multivariate generators.}
Second, further research 
may also explore and enhance our 
algorithm.
As the Quadratic Assignment Problem can be solved exactly in polynomial time in certain specific cases, {a}  
key area of 
{future work may be} the selection of an appropriate functional space, or sieve space, which has the potential to lead to more efficient computational algorithms. 
Third, while nonparametric inference offers flexibility, it is contingent on certain assumptions that may not always be fulfilled in real applications, and it is often {notably} 
impacted by the presence of a small number of outliers. 
This may {affect} 
the accuracy of the estimator but also the efficiency of the algorithm. {One way to deal with this issue is to couple SQL with robust} 
statistical methods (see e.g. \cite{hampel2011robust}, \cite{ronchetti2009robust}, and \cite{maronna2019robust} ){, which}  are designed to provide stable inference when slight deviations from the stochastic assumptions
on the model occur.
Robustification of the SQL approach may be achieved {using robust sieve M-estimators, as in}  
\cite{bodelet2022robust}.
{Additionally, we} conjecture that 
combining our results with Lemma 1 in \cite{bodelet2022robust} {may give} 
similar rates of convergence for the robustified SQL estimates.

To summarize, in the present study, we propose SQL, a large, nonlinear, and additive model suitable for the analysis of nonlinear high-dimensional data. The method leverages the flexibility of nonparametric statistics. Its theoretical properties allow one to quantify and assess the model performance given different sample sizes and feature dimensionality. Its applications to high-dimensional genetic data suggest its utility in both unsupervised learning (e.g., separating samples from different groups) and supervised learning (e.g., extracting latent features to predict disease outcomes). The enclosed SQL package (\url{https://github.com/jbodelet/SQL}) 
{helps} users to further explore and test the method and theory in a broad range of applications to investigate the nonlinear patterns of high-dimensional brain imaging data, gene expression data, and whole-body immunological biomarkers.

\appendix

\section{Proofs for the asymptotic theory}

\subsection{Notation}

We define $a\vee b = \max(a,b)$ and $a\wedge b = \min(a,b)$.
Throughout the proofs, we let $C$ be a generic absolute constant, whose value may change from line to line.
For two sequences $\alpha_n$ and $\beta_n$, we use $\alpha_n = o(\beta_n)$ to denote $\alpha_n / \beta_n \rightarrow 0$
and $\alpha_n = \bigO(\beta_n)$ to denote $\alpha_n \leq C\beta_n$ for all $n$ large enough.
Moreover, for a random sequence $X_n$, we use $X_n = o_p(a_n)$ to denote $X_n / a_n \rightarrow 0$ in probability, and
$X_n =\bigOp(a_n)$ if 
$\lim_{M \rightarrow \infty} \limsup_{n\rightarrow \infty}P(|X_n|>M \alpha_n) = 0$.

\subsection{Entropy
lemma}

Before considering the rates of convergence, we first provide a general bound on the entropy of the parameter space.
Given a space $\mathcal{G}_n$ and the space of permutations $\Pi_n$, we build the following space
$$
\mathcal{H}_n := \left\{h:\{1,\dots,n\}\times \{1,\dots,p\}\rightarrow \R, h_{\iov} = \sumfac g_\ivl\left( \frac{\pi^{(l)}_i}{n+1}\right), g_1,\dots,g_p \in \mathcal{G}_n^{\oplus q}, \pi^{(1)}, \dots, \pi^{(q)} \in \Pi_n\right\}.
$$
We consider the norm $\Vert h\Vert^2 := \frac{1}{np}\sumvar\sumobs h_\iov^2$.
The 
penalized estimator $\hat{h}$ can be then expressed as
$$
\hat{h} \in \argmin_{h \in \mathcal{H}_n} \Vert X-h\Vert^2 + \lambda I^2(h)
$$
where for $h_\iov = \sumfac g_\ivl(\frac{\pi^{(l)}}{n+1})$ we define with a slight abuse of notation $$
I^2(h):=I^2(\bm g)=\frac{1}{p}\sumvar \sumfac\int_0^1 |g_{j,l}^{(m)}(\xi)|^2 d\xi.
$$

We will also denote the true solution by $h^0_\iov:=\sumfac g_\ivl^0(Z_\iol^0)$ that satisfies
$$
X_\iov = h^0(i,j) + \epsilon_\iov.
$$
Note that $h^0$ may not belong to $\mathcal{H}_n$.
We then define 
$$
h_n^* \in \argmin_{h \in \mathcal{H}_n} \Vert h-h^0\Vert^2 + \lambda I^2(h),
$$
where $h_n^*$ can be interpreted as the ``closest'' element in $\mathcal{H}_n$ to $h^0$.
We also denote by $\bm g^*$ and $\bm\pi^*$ the functions and permutation vector that satisfy
$h^*_\iov = \sumfac g^*_\ivl(\frac{\pi^{*(l)}}{n+1})$.

We also define, for any $\delta>0$, the family of sets
$$\mathcal{H}_n(\delta):= \{h \in \mathcal{H}_n,\; \Vert h-h^*\Vert^2 + \lambda I^2(h) \leq \delta^2 \}.$$
Moreover, for $g \in \mathcal{G}_n^{\oplus q}$, we let the empirical norm be 
$$
\Vert g \Vert_n^2:= \frac{1}{n}\sumobs \sumfac\left(g_\ivl\left(\frac{i}{n+1}\right)\right)^2,$$
and we define, for any $\delta>0$, the family of sets
$$
\mathcal{G}_n^{\oplus q}(\delta):= \{g \in \mathcal{G}_n^{\oplus q}, \; \Vert g-g^*\Vert^2_n + \lambda I^2(g) \leq \delta^2 \}.
$$
For a set $\mathcal{S}$ and norm $\Vert \cdot\Vert$, we define the covering number $N(\varepsilon, S, \Vert \cdot\Vert)$ to be the smallest number of balls of radius $\delta$ to cover $\mathcal{S}$.
We call $H(\varepsilon, S, \Vert \cdot\Vert) := \log N(\varepsilon, S, \Vert \cdot\Vert)$ the entropy of $\mathcal{S}$ and we define the entropy integral as
$$
J(\delta, S, \Vert \cdot\Vert)=
\int_{0}^\delta H^{1/2}(\varepsilon, \mathcal{S}, \Vert \cdot \Vert) d\varepsilon,
$$
provided it exists.

The following lemma states that the metric entropy of $\mathcal{H}_n(\delta)$ can be computed from the metric entropy of $\mathcal{G}_n(\delta)$.

\begin{lemma}\label{th. lemma entropy}
Let Assumption \ref{ass. rates mixing} and \ref{ass. rates Lipschitz} be satisfied and
suppose that, for any $\varepsilon>0$ we have $N(\varepsilon, \mathcal{G}_n(\delta/q), \Vert \cdot \Vert_n)< \infty$.
We have for some $C>0$ that
$$H(\varepsilon,\mathcal{H}_n(\delta), \Vert \cdot \Vert)
\leq q \log(\no!)
+
pqH(\varepsilon,\mathcal{G}_n(\delta/q), \Vert \cdot \Vert_n)$$
for all $\varepsilon > 0$ and any $\lambda \geq 0$. Moreover, we have
$$
J(\delta, \mathcal{H}_n(\delta), \Vert \cdot \Vert) \leq \sqrt{q \log(n!)} \delta + \sqrt{pq} J(\delta, \mathcal{G}_n(\delta/q), \Vert \cdot \Vert_n).
$$
\end{lemma}

\subsection{Proof of Lemma \ref{lemma. approximation error}}

First, define $(\bm g^*,\bm\pi^*)$ as minimizers of 
$$
\totalmean
\left( 
\sumfac \Rfunc_\ivl\left(\frac{\pi_\io^{(l)}}{n+1}\right) -\func_\ivl^0(\Factor_\io^0) \right)^2
+ \lambda I^2(\bm g).
$$
Using the inequality $(a+b)^2\leq 2a^2+2b^2$ we obtain
\begin{align*}
&\totalmean
\left( 
\sumfac \Rfunc_\ivl^*\left(\frac{\pi_\io^{*(l)}}{n+1}\right) -\func_\ivl^0(\Factor_\io^0) \right)^2
+ \lambda I^2(\bm g^*)
\\
&\leq \min_{g_j \in \mathcal{G}_n^{\oplus q}} \min_{\pi^{(l)}\in \Pi_n}
\totalmean
\bigg( 
\sumfac \Rfunc_\ivl^*\left(\frac{\pi_\io^{*(l)}}{n+1}\right) -\func_\ivl\circ P_Z^{-1}\left(\frac{\pi_\io^{(l)}}{n+1}\right) 
+ \func_\ivl\circ P_Z^{-1}\left(\frac{\pi_\io^{(l)}}{n+1}\right) 
-\func_\ivl^0(\Factor_\io^0) \Bigg)^2    + \lambda I^2(\bm g)\\
&\leq \min_{g_j \in \mathcal{G}_n^{\oplus q}} \frac{2}{\nov}\sumobs \sumvar
\left( \sumfac \Rfunc_\ivl\left(\frac{\pi_\io^{(l)}}{n+1}\right) -
\func_\ivl\circ P_Z^{-1}\left(\frac{\pi_\io^{(l)}}{n+1}\right) 
\right)^2
 + \lambda I^2(\bm g)+ \\
&\min_{\pi^{(l)}\in \Pi_n}
\frac{2}{\nov}\sumobs \sumvar
\left( 
\sumfac\func_\ivl^0\circ P_Z^{-1}\left(\frac{\pi^{(l)}_\io}{n+1} \right)
-\func_\ivl^0(\factor_\iol^0)  \right)^2.
\end{align*}

For the first component, we obtain
\begin{equation*}
\min_{g_j \in \mathcal{G}_n^{\oplus q}} \frac{2}{\nov}\sumobs \sumvar
\left( \sumfac \Rfunc_\ivl\left(\frac{\pi_\io^{(l)}}{n+1}\right) -
\func_\ivl\circ P_Z^{-1}\left(\frac{\pi_\io^{(l)}}{n+1}\right) 
\right)^2  + \lambda I^2(\bm g) 
\leq 2\tau_n^*
\end{equation*}
where 
$$
\tau_n^* = \min_{g_j \in \mathcal{G}_n^{\oplus q}} \frac{1}{\nov}\sumobs \sumvar
\sumfac  \left(\Rfunc_\ivl\left(\xi_i\right) -
\func_\ivl\circ P_Z^{-1}\left(\xi_i\right) 
\right)^2     + \lambda I^2(\bm g)
$$

For the second component, we apply Assumption \ref{ass. rates Lipschitz} to get that, for any $\factor_\iol$,
\begin{align*}
\frac{2}{\nov}\sumobs \sumvar
         \left(\sumfac \func_{\ivl}^0\left(\frac{\pi^{(l)}_\io}{n+1} \right) -  \func_{\ivl}^0( \factor_\iol^0) \right)^2
&\leq
\frac{2q}{\nov}\sumobs\sumvar \sumfac 
         \left(\func_{\ivl}^0\left(\frac{\pi^{(l)}_\io}{n+1} \right) -  \func^0_{\ivl}( \factor_\iol^0) \right)^2\\
&\leq
\frac{2L_n^2\nf}{\no}\sumobs \sumfac \left(P_Z^{-1}\left(\frac{\pi^{(l)}_\io}{n+1}\right)  -  \factor_\iol^0\right)^{ 2}.
\end{align*}
Now we will show that, when taking the minimum over $\bm \pi$, the right hand side of the above inequality converges
with rate $n^{-1}$.
Denote the $k$-th order statistics by $\factor^0_{(k),l}$, so we have  $\factor^0_{(1),l} \leq \factor^0_{(2),l}\leq \dots \leq \factor^0_{(n),l}$.
Note that we can write:
$$
\min_{ \pi^{(l)}}\frac{1}{n}\sumobs \left(P_Z^{-1}\left(\frac{\pi^{(l)}_\io}{n+1}\right)  -  \factor_\iol^0\right)^{ 2}
= 
\meanobs \left( P_Z^{-1}\left(\frac{\io}{\no+1}\right) -  \factor_{(\io),l}^0\right)^{2}.
$$
Denote by $P_{\no}^{(l)}(z):=\frac{1}{\no+1}\sum_{\io = 1}^\no \mathbbm{1}(\factor_{\iol} \leq z )$ the empirical distribution of the $l$-th factors, and
by $P^{(l)-1}_{\no,l}(p) := 
\inf \{z: P_{\no}^{(l)}(z)\geq p\}
$ the corresponding empirical quantile function.
Given Assumption \ref{ass. rates mixing}, we can use the results in 
Lemma 21.4 and Corollary 21.5 in \citet{vandervaart2000asymptotic} which shows that the empirical quantile function converges to the theoretical
quantile function with the same rate of converge as the cumulative distribution function.
In particular, we have, for any $\xi \in (0,1)$, that
$$
\sqrt{\no} \left(P_{\no}^{(l)-1}(\xi) -  P^{-1}_Z(\xi) \right)= \frac{-1}{\sqrt{n}}\sum_{\io = 1}^\no \frac{\mathbbm{1}(\factor_{\iol} \leq P_Z^{-1}(\xi) )-  P_Z(\xi) }{p_Z^{(l)}(P_Z^{-1}(\xi))}+o_p(1).
$$
As we have
$\factor_{(\io),l}^0 = P_{\no}^{(l)-1}(\frac{\io}{\no+1})$, we can apply the egodic property in Assumption \ref{ass. rates mixing} to obtain for some constant $K>0$ that
$$ \min_{ \pi^{(l)} } \meanobs \left( \frac{\pi_i^{(l)}}{n+1} -  \factor_\iol^0\right)^{ 2} \leq
K\sup_z
\left|P_{\no}^{(l)}(z) -  P_Z(z) \right|^2 +o_p\left(\frac{1}{n}\right)
= O_p\left(\frac{1}{n}\right).$$
Combining the above results gives
$$
\min_{g_j \in \mathcal{G}^{\oplus q}, \pi^{(l)}\in \Pi_n}\totalmean \left( \sumfac\Rfunc_\ivl\left( \frac{\pi_i^{(l)}}{n+1}\right) -  \func_\ivl^0( \factor_\iol^0) \right)^2
= \bigOp\left(\tau_n^* + \frac{L_n^2}{n}  \right).
$$

\subsection{Proof of Lemma \ref{th. lemma entropy}}

Let $\mathcal{C}_g:=\{g^1,\dots,g^N\}$ denotes the $\varepsilon$-covering of $\mathcal{G}^{\oplus q}(\delta)$
with $N=N(\varepsilon, \mathcal{G}^{\oplus q}(\delta), \Vert \cdot \Vert_n)$.
Let's choose $\tilde h \in \mathcal{H}_n(\delta)$, defined by $\tilde h(i,j)= \sumfac \tilde g_\ivl\left(\frac{\tilde \pi^{(l)}}{n+1}\right)$
where $\tilde g_j \in \mathcal{G}^{\oplus q}(\delta)$.
There are $k_1, \dots, k_p$ such that 
$g^{k_j} \in \mathcal{C}_g$ with 
$\Vert \tilde g_j- g^{k_j}\Vert_n\leq \varepsilon$. We can then build $h$ that satisfies 
$ h(i,j)= \sumfac g_l^{k_j}\left(\frac{\tilde \pi^{(l)}}{n+1}\right)$.
Obvisously we have $\Vert \tilde h - h\Vert \leq 
\sqrt{\meanvar \Vert \tilde g_j - g^{k_j}\Vert^2_n}\leq \varepsilon $.
The set 
$$
\mathcal{C}_h = \left\{h, h(i,j) = \sumfac g_\il^{k_j}\left( \frac{\pi^{(l)}_i}{n+1}\right), g^{k_1},\dots,g^{k_p} \in \mathcal{C}_g, \pi^{(1)}, \dots, \pi^{(q)} \in \Pi_n\right\}.
$$
is thus an $\varepsilon$-covering of $\mathcal{H}_n(\delta)$ with respect to $\Vert \cdot \Vert$. Its cardinal $\mathcal{C}_h$ is given obviously by $(n!)^qN^p$.
We conclude that 
\begin{equation}\label{eq. app. bound on entropy}
H(\varepsilon,\mathcal{H}_n(\delta), \Vert \cdot \Vert) = \log N(\varepsilon,\mathcal{H}_n(\delta), \Vert \cdot \Vert)
\leq q\log(\no!)
+ p H(\varepsilon,\mathcal{G}_n^{\oplus q}(\delta), \Vert \cdot \Vert_n).    
\end{equation}
Now we use Lemma 8 in \cite{sadhanala2019additive} which computes bounds on additive sets to get
$$
H(\varepsilon,\mathcal{G}_n^{\oplus q}(\delta), \Vert \cdot \Vert_n)
\leq 
q H(\varepsilon,\mathcal{G}_n(\delta/q), \Vert \cdot \Vert_n)
$$
Taking the integral of the square root of the right hand side of \eqref{eq. app. bound on entropy} gives
a bound on Dudley's integral:
$$
J(\delta, \mathcal{H}_n(\delta), \Vert \cdot \Vert) \leq \sqrt{q \log(n!)} \delta + \sqrt{pq} J(\delta, \mathcal{G}(\delta), \Vert \cdot \Vert_n).
$$

\subsection{Proof of Theorem \ref{th. rates of convergence}}

The total error is separated into an approximation and the estimation error.
Using $g_j^*$ and $\pi^*$ in Lemma \ref{lemma. approximation error}, we  decompose the total error into an estimation error and an approximation error as follows:
\begin{align*}
\sqrt{\totalmean
\left( 
\hat \func_\iv(\hat \Factor_\io)-\func_\iv^0(\Factor_\io^0)  \right)^2}
&\leq \\
\underbrace{\sqrt{\totalmean
\left( \sumfac \hat \func_\ivl(\hat \factor_\iol)-
\Rfunc_\ivl^*\left(\frac{\pi_i^{*(l)}}{n+1}\right)  \right)^2}
}_{\text{estimation error}}
&+
\underbrace{\sqrt{\totalmean
\left( \sumfac
\Rfunc_\ivl^*\left(\frac{\pi_i^{*(l)}}{n+1}\right) 
-\func_\ivl^0(\factor_\iol^0)  \right)^2}
}_{\text{approximation error}}.
\end{align*}

The approximation error is bounded using Lemma \ref{lemma. approximation error}. 
From Assumption \ref{ass. rates splines} we have
\begin{equation}\label{eq. sieve func error}
\tau^*_n = \bigO(d^{-2\eta}).
\end{equation}

Now we just need a bound for the estimation error.
Let $\Psi(\delta) \geq J(\delta, \mathcal{H}_n(\delta), \Vert \cdot \Vert)$ be such that $\Psi(\delta)/\delta^2$ is a decreasing function of $\delta$.
Applying Theorem 10.11 in \cite{VandeGeer2000}, which treats rates of convergence for least squares estimators on sieves, for some $c$ depending on $K$ and $\sigma$ and for $\delta_\nov$ satisfying 
\begin{equation}\label{eq. proof Cor critical inequality}
\sqrt{np} \delta_\nov^2 \geq c \Psi(\delta_\nov),\quad \text{ for all } n,
\end{equation}
we have
\begin{equation}\label{eq. proof cor probatility bound}
\norm{\hat \compFunc- \compFunc^*}= \bigO(\delta_\nov). 
\end{equation}

We use Lemma \ref{th. lemma entropy} to compute the entropy. 
The $\varepsilon$-covering number of $\mathcal{G}_n(\delta)$ can be  bounded by
$$N(\varepsilon, \mathcal{G}_n(\delta/q), \Vert \cdot \Vert_n) = \bigO\left( \left(\frac{4  \delta/q + \varepsilon}{\varepsilon}\right)^{d} \right),$$
where we used Corollary 2.6 in \citet{VandeGeer2000}, which states that the $\epsilon$-covering number for functions $g(\cdot) = \sumbasis b_k \psi_k(\cdot)$ such that $\Vert\sumbasis b_k \psi_k(\cdot)\Vert_n \leq M$
is of order $((4 M + \varepsilon)/\varepsilon)^d$.
We then get
\begin{equation}
H(\varepsilon, \svt(\delta), \Vert \cdot \Vert)
= \bigO\left(
\log(\no!)  + pq d \log\left(\frac{4  \delta/q + \varepsilon}{\varepsilon}\right)
\right).
\end{equation}

Integrating the square root of the entropy bound then 
yields
\begin{align*}
J(\delta, \svt(\delta), \Vert \cdot \Vert)
&\leq  C\delta  \left[ \sqrt{\log \no!} + \sqrt{d \nv} \int_0^1 \sqrt{\log(4 /q + v)} dv
\right],\\
&\leq \tilde C \delta \left( \sqrt{\log \no!} + \sqrt{d \nv} \right),
\end{align*}
for some positive constants $C$ and $\tilde C$ which depend on $q$.
Using inequality \eqref{eq. proof Cor critical inequality}, it follows that \eqref{eq. proof cor probatility bound} is satisfied for any $\delta_\nov$ such that, for some $\tilde c>0$,
$$\delta_\nov^2 \geq \tilde c \left( \frac{\log \no}{\nv} + \frac{d}{\no}
\right),$$
where Stirling formula gave $\log \no! \asymp \no \log \no$.
The theorem follows by combining \eqref{eq. proof cor probatility bound} with Lemma \ref{lemma. approximation error} and \eqref{eq. sieve func error}.

\subsection{Proof of Theorem \ref{th. rates of convergence penalyzed}}

We first state the following lemma.
\begin{lemma}\label{lemma. rates penalyzation}
    Suppose Assumptions \ref{ass. rates mixing}, \ref{ass. rates Lipschitz}, and \ref{ass. rates subgaussian} are met. Take  $\Psi(\delta) \geq J(\delta, \mathcal{H}_n^*(\delta), \Vert \cdot \Vert_\nov)$ such that $\Psi(\delta)/ \delta^2$ is a decreasing function of $\delta$, $0<\delta<2^7\sigma_0$.
    Then for 
    $$
    \sqrt{\nov}\delta_\nov^2 = \Psi(\delta_\nov), \forall n, p
    $$
    we have
    $$
    \Vert \hat{h}-h^0\Vert^2 
+ \lambda I^2(\hat{h})
= \bigOp\left(\delta_\nov^2 +\Vert h^*-h^0\Vert^2
+\lambda I^2(h^*) \right).
$$
\end{lemma}
The proof of Lemma \ref{lemma. rates penalyzation} will be shown later.
To conclude the theorem, we need to compute $\delta_\nov$ and the approximation error $\tau_n^*$.

We first compute the entropy for the penalized estimators.
We use Lemma \ref{th. lemma entropy} and consider $\mathcal{G}_n= \mathcal{G}$ as the Sobolev space of $m$-th continuously differentiable functions.
From \cite{birman1967piecewise}, we have that
$$
H\left(\delta, \left\{g[0,1]\rightarrow \R, \Vert g\Vert_\infty \leq 1, \int_0^1 g^{(m)}(\xi)d\xi \leq 1\right\}, \Vert \cdot \Vert_\infty\right)
\leq C \delta^{-1/m}, \quad \delta >0.$$
From this, following the same steps as in \cite{vandegeer2000empirical}, we can show that
$$
H(\delta, \mathcal{G}(\delta), \Vert \cdot \Vert_n) \leq C\left(
\delta^{1/m} \lambda^{-1/2m} \varepsilon^{-1/m} + \log\left(\frac{\delta}{(\sqrt{\lambda} \wedge 1) \varepsilon}\right)
\right), \quad 0<\varepsilon<\delta,
$$
which gives
$$
J(\delta, \mathcal{G}(\delta), \Vert \cdot \Vert_n) \leq \tilde C\left(
\delta \lambda^{-1/2m} + \delta \sqrt{\log(1/\sqrt{\lambda} \vee 1)}
\right), \forall \delta > 0,
$$
where $\tilde C$ depends on $m$.
Using Lemma \ref{th. lemma entropy} we get
$$
J(\delta, \mathcal{H}(\delta), \Vert \cdot \Vert) \leq C\delta \left(
\sqrt{\log(n!) }+ 
\sqrt{p}\lambda^{-1/2m} + 
\sqrt{p\log(1/\sqrt{\lambda} \vee 1)}
\right), \forall \delta > 0.
$$
By Lemma \ref{lemma. rates penalyzation} we can select $\delta_{np}$ satisfying
$$
\sqrt{np}\delta_{n,p} \geq \left(
\sqrt{\log(n!) }+ 
\sqrt{p}\lambda^{-1/(2m)} + \sqrt{p\log(1/\sqrt{\lambda} \vee 1 )}
\right),
$$
which gives
$$
\delta_{n,p}^2 \geq 
C\left(
p^{-1}\log(n)+ 
n^{-1}\lambda^{-1/(2m)}+ 
n^{-1}\log(1/\sqrt{\lambda} \vee 1 )
\right).
$$

Now we bound the approximation error.
From Lemma \ref{lemma. approximation error} we have
$$
\Vert h^*-h^0\Vert^2
+\lambda I^2(h^*) = \bigOp\left(\frac{L_n^2}{n}+ \tau_n^*\right).
$$
We then use $\tau^*_n \leq \lambda I^2(\bm g^0)$ to get
$$
\Vert h^*-h^0\Vert^2
+\lambda I^2(h^*) = \bigOp\left(\frac{L_n^2}{n}+ \lambda I^2(\bm g^0)\right).
$$
Combining the bounds for the approximation and estimation errors concludes the proof.

\subsection{Proof of Lemma \ref{lemma. rates penalyzation}}

The estimator $\hat{h}$ satisfies
$$
\Vert X-\hat{h}\Vert^2 + \lambda I^2(\hat{h})
\leq \Vert X-h^*\Vert^2
+\lambda I^2(h^*),
$$
and thus
$$
\Vert \hat{h}-h^0\Vert^2 + \lambda I^2(\hat{h})
\leq 
2\sup_{h\in \mathcal{H}_n} 
\langle \epsilon, \hat{h}-h^*\rangle
+ 
\Vert h^*-h^0\Vert^2
+\lambda I^2(h^*)
$$
where we used the notation $\langle x,y\rangle = \frac{1}{np} \sumobs\sumvar x_\iov y_\iov$
for the empirical inner product.
Moreover, we have
$$
\Vert \hat{h}-h^*\Vert^2 \leq
2\left( \Vert \hat{h}-h^0\Vert^2 + 
\Vert h^*-h^0\Vert^2 \right).
$$
We then add $\lambda I^2(\hat{h})$ on both sides of the inequality to get 
\begin{equation*}
\Vert \hat{h}-h^*\Vert^2 
+ \lambda I^2(\hat{h})
\leq
2\Vert \hat{h}-h^0\Vert^2 + 
2\Vert h^*-h^0\Vert^2
+ \lambda I^2(\hat h)
\end{equation*}
and use the above supremum bound to get the basic inequality
\begin{equation}\label{eq. Pf Lemma3 basic inequality}
\Vert \hat{h}-h^*\Vert^2 
+ \lambda I^2(\hat{h})
 \leq 
4\sup_{h\in \mathcal{H}_n} 
\langle\epsilon, \hat{h}-h^*\rangle
+ 
4\Vert h^*-h^0\Vert^2
+2\lambda I^2(h^*). 
\end{equation}
On the set where $\langle\epsilon, \hat{h}-h^*\rangle <
\Vert h^*-h^0\Vert^2
+\lambda I^2(h^*)
$
the Lemma holds.
On the set where 
$\langle\epsilon, \hat{h}-h^*\rangle \geq
\Vert h^*-h^0\Vert^2
+\lambda I^2(h^*)
$
we have 
$$
\Vert \hat{h}-h^*\Vert^2 
+ \lambda I^2(\hat{h})
 \leq 
8\sup_{h\in \mathcal{H}_n} 
\langle\epsilon, \hat{h}-h^*\rangle.
$$
We can now apply the same arguments as in Theorem 9.1 in \cite{VandeGeer2000}.

We then conclude by applying the triangle inequality
$\Vert \hat{h}-h^0\Vert \leq 
\Vert \hat{h}-h^*\Vert +
\Vert h^*-h^0\Vert
$.

\section{Proof of Theorem \ref{th. identification}}

First we show that there exists an isomorphism $\bm H:\R^\nf \rightarrow \R^\nf$ such that $\Factor= \bm H(\tilde \Factor)$.
Assumption \ref{ass. identification sufficient nonlinearity} implies that the Jacobian of $\bm f$ is full rank and hence $\bm f$ is injective.
In particular, $\bm f$ is a diffeomorphism onto its image, $\bm f(\mathcal{Z}^q)$.
Then we show that $\tilde{\bm f}$ has to be a diffeomorphism on the same image $\bm f(\mathcal{Z}^q)$.
As a bijection (i.e., $\bm f$) exists between these two sets, they have same size.
$\tilde{\bm f}$ has the same image and same support and thus must be a bijection too.
Now for any $\bm z\in \R^q$, $\exists!\bm x\in \R^p$ such that $\bm x=\bm f(\bm z)$. 
Then it exists a unique $\tilde{\bm z}\in \R^q$ such that $\bm x=\tilde{\bm f}(\tilde{\bm z})$. We then denote by $\bm H$ the invertible map that links this $\bm z$ to this $\tilde{\bm z}$.
We have then
$$
\tilde{\bm \Func}(\tilde{\bm Z}) = \Func(\Factor)= \Func(\bm H(\tilde{\bm Z}))
$$
and thus $\tilde \func_\iv(\cdot) = \func_\iv \circ \bm H(\cdot) $.
Let $H_1(\cdot), ..., H_\nf(\cdot)$ be the entries of $\bm H$.
In order $\tilde \func_\iv(\cdot)$ to satisfy Assumption \ref{ass. identification function ordering} $\bm H$ has to be twice differentiable.
Moreover, since $\tilde \func_j$ are additive, it holds that for any $k',k \in \{1,\ldots,\nf\}$ and $ k' \neq k$ 
$$ \frac{\partial^2}{\partial z_k z_{k'}} \tilde g_\iv(\bm z) = 0, \text{for } k \in \{1,\ldots,\nv\}$$
for any $\bm z = (z_1, \ldots, z_\nf)^\top \in \R^q$.
Since $\tilde \func_\iv = \func_\iv \circ \bm H$, it yields, for any $k' \neq k$, that
\begin{equation}\label{eq. deriv of g tilde proof th 1}
\sumfac g_\ivl'(H_\il(\bm z))
\frac{\partial^2 H_\il(\bm z)}{\partial z_k \partial z_{k'}}  + g_\ivl''(H_\il(\bm z))
\frac{\partial H_\il(\bm z)}{\partial z_k}  \frac{\partial H_\il(\bm z)}{\partial z_{k'}} = 0, \text{for } j \in \{1,\ldots,\nv\}.
\end{equation}
For any $\bm z \in \R^\nf$ and $k\neq j$, we thus have the linear systems of equations,
$$ \bm \Delta(\bm H(\bm z)) \bm a_{kk'}(\bm z)=0,$$ 
where $\bm \Delta(\bm z)$ is the $\nv \times 2\nf$-dimensional matrix with rows
$(\func_{\iv 1}'(z_1), \ldots, \func_{\iv \nf}'(z_\nf), \func_{\iv 1}''(z_1), \ldots, \func_{\iv \nf}''(z_\nf))$
and we define
$$ \bm a_{kk'}(z) = \left(
\frac{\partial^2 H_1(\bm z)}{\partial z_k \partial z_{k'}}, \ldots,
\frac{\partial^2 H_\nf(\bm z)}{\partial z_k \partial z_{k'}},
\frac{\partial H_1(\bm z)}{\partial z_k}  \frac{\partial H_1(\bm z)}{\partial z_{k'}}, \ldots, \frac{\partial H_\nv(\bm z)}{\partial z_k}  \frac{\partial H_\nv(\bm z)}{\partial z_{k'}} \right)^\top.$$
We now apply Assumption \ref{ass. identification sufficient nonlinearity} which specifies that $\bm \Delta(\bm z)$ is full rank for any $\bm z \in \R^\nf$.
This implies that there is only one unique solution $\bm a_{kk'}(\bm z)= 0$ for any $k\neq k'$ and thus
$$\frac{\partial^2 H_\il(\bm z)}{\partial z_k \partial z_{k'}} = 0 \quad \text{and} \quad
\frac{\partial H_\il(\bm z)}{\partial z_k}  \frac{\partial H_\il(\bm z)}{\partial z_{k'}} = 0,$$
for any $\setl$.
This implies that each entry
$H_\il(\bm z)$ depends on at most one of the $z_1, ..., z_\nf$, i.e., each $H_\il(\cdot)$ is a univariate function of one of the element of $(z_1, \dots, z_q)$.
We can thus write by $H_\il(\bm z) = h_\il(z_{\sigma(\il)})$ for a univariate function $h_\il$ and $\sigma_l \in \{1, \dots, q\}$ for  $l\in \{1,\ldots,\nf\}$.
We thus have:
$$ \tilde g_\iv(\bm z) = \sumfac g_\ivl(h_\il (z_{\sigma_\il})),
$$
By applying $\tilde g_j$ to $\tilde \Factor$ we obtain $\tilde g_\iv(\tilde \Factor) = \sumfac g_\ivl(h_\il (\tilde \factor_{\sigma_\il}))$ and thus
$$\factor_\il = h_\il(\tilde\factor_{\sigma_l}).
$$
As we assumed that
$\tilde \factor_\iol$ and $\factor_{\io,\sigma_l}$ are both standard normally distributed variables.
As $h_\il$ have to be smooth, 
$\tilde \factor_\il = \factor_{\il'}$ or $-\factor_{\il'}$ are the only two solutions.
In order to not contradict Assumption \ref{ass. identification intrinsic dimension} $\sigma_l$ has to be bijective, and thus a permutation.
Finally, because in Assumption \ref{ass. identification function ordering} we assumed that there is a unique ordering as given by the $L_2$ norm of the functions $g_\ivl$, we have $\il'=\il$. This completes the proof.

\section{ Implementation of the VAE}
\label{apx:vae-implementation}

A Variational Autoencoder (VAE), introduced by \cite{Kingma2014} allows the flexible modeling of multivariate data $\bm X$ by specifying their distribution conditional on latent variables $\bm z \in \mathcal Z$. Its objective is twofold: the first is to estimate the generative model $p_{\bm\theta}(\bm X|\bm z)$, corresponding to the density of $\bm X$ conditional on $\bm z$, which corresponds to a neural network parameterized by $\bm\theta$ (the decoder). The second is to approximate the posterior density of the latent variables given the observed data with a neural network $q_{\bm\phi}(\bm z|\bm X)$ parameterized by $\bm\phi$ (the encoder).

The estimators $(\hat{\bm\phi}, \hat{\bm\theta})$  of $(\phi, \theta)$ are obtained by minimizing the following loss function:
$$
\mathcal{L}(\phi, \theta; {\bm X}) = -\mathbb{E}_{q_{\phi}(\mathbf{z}|{\bm X})} [\log p_{\theta}({\bm X}|\mathbf{z})] + \mathrm{KL}(q_{\phi}(\mathbf{z}|{\bm X}) \ || \ p(\mathbf{z})),
$$
where $\mathrm{KL}$ is the Kullback-Leibler divergence between the distribution produced by the encoder $q_{\phi}(\mathbf{z}|{\bm X})$ and some prior distribution $p(\mathbf{z})$, which, according to our model, is standard multivariate Gaussian. We follow the common practice of modeling the encoder's output distribution $q_{\bm \phi}$  as a multivariate normal distribution $\mathcal{N}(\bm\mu, \bm\Sigma)$, where $\bm\mu$ and $\bm\Sigma$ are the mean vector and covariance matrix outputted by the encoder neural network.

The first part of the loss function can be seen as an approximation of the standard Expectation-Maximization step \citep{dempster1977maximum}, where the true posterior distribution of $\bm z|\bm X$ is replaced by $q_{\bm\phi}(\bm z|\bm X)$; the second is a regularization term that attempts to correct for the approximation. Under some regularity conditions it can be shown that $\mathcal{L}(\phi, \theta; {\bm X})$ is a lower bound for the true (marginal) likelihood of the data $\mathcal L(\xi; \bm X)$, given by
$$
\mathcal L(\xi; \bm X) = \int_{{\mathcal{Z}^q}}\left(\bm \mu + \sum_{l=1}^q g_{j,l}(z_{l})\right)dP_{Z}(\bm z),
$$
with $\bm z = (z_1, \dots, z_q)$ (see \oyc{e.g.,} \citealp{Kingma2014}).

A VAE is a flexible model, comprising of two neural networks. This increased flexibility, however, makes interpretation challenging and is prey to overfitting. In our case, given our knowledge of the true generative model in Equation \eqref{eq. additive model}, it is natural in the decoder to model each function $g_{j,l}$ independently by a feed-forward neural network denoted by $\tilde g_{j,l}$. Specifically, the decoder's task is to first estimate pseudo-basis functions, which are then linearly combined to produce the estimated $\tilde g_{j,l}$ functions for all $j$. Imposing this structure not only allows the estimation of these functions, themselves of interest, but also makes for a more parsimonious model, potentially making for a better estimation of the latent variable $z$. What's more, the strategy mimmics the SQL, which, in our view, provides a fair comparison. To avoid overfitting, regularization techniques such as dropout or L2 regularization can be employed \citep{srivastava2014dropout}. A good overview of feed-forward neural network and regularization and other optimization techniques can be found in \citep{goodfellow2016deep}. Additionally to using dropout, we consider another strategy in which the outermost layer is constrained to have just $L'$ neurons: this choice mirrors the basis functions modeling of our paper, in the sense that the neural network learns to map the latent space to each of the $L'$ neurons, which can thus be loosely interpreted as estimated basis functions (albeit without any orthogonality constraints), and which are then linearly combined to obtain the fitted data outputted by the decoder. We select  $L'$ such as to reduce the out-of-sample prediction error.

We now present in greater details the chosen architecture for the VAE. 

\subsubsection*{Encoder}
\begin{itemize}
  \item The input is the observed data $\bm X$.
  \item The data then passes through a dense layer with 100 neurons and tanh activation.
  \item For both the mean ($\bm\mu$) and log-variance (the log of the diagonal elements of $\bm\Sigma$) of the latent space distribution:
  \begin{itemize}
    \item Two dense layers with 50 neurons each, using tanh activation.
    \item A final dense layer with linear activation, with output dimension the same as that of the latent space.
  \end{itemize}
\end{itemize}

\subsubsection*{Decoder}
\begin{itemize}
  \item The input is the latent variable $z$.
  \item A dense layer with 100 neurons and tanh activation processes $z$.
      \item Two dense layers with 50 neurons each, tanh activation.
      \item A penultimate layer with $L'$ neurons, tanh activation.
      \item A final layer outputs the reconstructed data.
\end{itemize}

The use of tanh activation functions is motivated by the a priori knowledge of smooth functions.

\bibliographystyle{apalike}
\bibliography{manual.bib}

\end{document}